\documentclass[a4paper,11pt]{article}
\pdfoutput=1
\usepackage{jheppub}
\usepackage[T1]{fontenc}
\usepackage{diagbox}
\usepackage{adjustbox}
\usepackage{multirow}
\usepackage{pifont}
\usepackage{array}
\usepackage{mathtools}
\usepackage{booktabs}
\usepackage[dvipsnames]{xcolor}
\usepackage{tcolorbox}
\usepackage{float}
\usepackage{placeins}
\usepackage{braket}
\usepackage{slashed}

\newcommand{\vev}{v_\text{EW}}

\usepackage[normalem]{ulem}
\newcommand{\stkout}[1]{\ifmmode\text{\sout{\ensuremath{#1}}}\else\sout{#1}\fi}

\newcommand{\beq}{\begin{equation}}
\newcommand{\eeq}{\end{equation}}
\newcommand{\bea}{\begin{eqnarray}}
\newcommand{\eea}{\end{eqnarray}}

\title{ 
A Non-linear Representation of General Scalar Extensions of the Standard Model for HEFT Matching
}

\author[a]{Huayang Song,}
\author[b]{Xia Wan,}

\affiliation[a]{Particle Theory and Cosmology Group, Center for Theoretical Physics of the Universe, Institute for Basic Science (IBS), Daejeon, 34126, Korea}
\affiliation[b]{School of physics and Information Technology, Shaanxi Normal University,
Xi'an 710119, China}

\emailAdd{huayangs1990@ibs.re.kr}
\emailAdd{wanxia@snnu.edu.cn}

\abstract{
We introduce a non-linear representation of ultraviolet~(UV) complete model, $U$ representation, under which matching HEFT to general scalar extensions of the standard model is straightforward. 
The main idea is to express a scalar multiplet in its linear form rotated by $U$ matrices, 
where $U$ matrix is exponential form of Goldstones based on Pauli matrices
and meanwhile is a special $SU(2)_L$ rotation. All together the doublet is expressed in $U$ matrix multiplying a doublet column or bidoublet matrix, which is composed of physical Higgs.
We show a complete matching between HEFT and real triplet extension. Meanwhile, under tensor notation, we give $U$ representation for general scalar extensions.
}

\begin{document} 
\maketitle
\flushbottom

\section{Introduction\label{section:introduction}}

After the discovery of the Higgs boson, the last piece of the Standard Model (SM), at the Large Hadron Collider (LHC) in 2012~\cite{CMS:2012qbp,ATLAS:2012yve}, the major efforts of the experiments (especially the LHC) has turned to searches for new particles at higher energy scales and precision measurements of Higgs and electroweak (EW) observables. So far no significant deviation from the SM predictions has been observed, which indicates that the potential new physics beyond the SM (BSM), if it exists, must be at a higher energy scale than that probed by the current experiments. In this scenario, effective field theories (EFTs) provide us the general and consistent framework to parameterize the BSM effects and possible deviations of the SM in the future experiments.

Two types of EFTs are commonly used to model the BSM physics in high-energy physics community - the SM Effective Field Theory (SMEFT)~\cite{Brivio:2017vri} and the Higgs Effective Field Theory (HEFT)~\cite{Appelquist:1980vg, Longhitano:1980iz, Longhitano:1980tm, Feruglio:1992wf, Herrero:1993nc, Herrero:1994iu, Grinstein:2007iv, Buchalla:2012qq, Alonso:2012px, Buchalla:2013rka, Brivio:2013pma, Buchalla:2013eza, Gavela:2014vra, Pich:2015kwa, Alonso:2015fsp, Brivio:2016fzo, Alonso:2016oah, Pich:2016lew, Merlo:2016prs, Pich:2018ltt, Krause:2018cwe, Sun:2022ssa, Sun:2022snw}. Both are constructed by exclusive SM degrees of freedom and invariantly under the SM gauge group $SU(3)_c\times SU(2)_L\times U(1)_Y$. However, they treat the Higgs field, $H_0$ and the electroweak Goldstone bosons, $\pi_i,i=1,2,3$ differently. The SMEFT embeds $H_0$ and $\pi_i$'s in an $SU(2)_L$ scalar doublet while the HEFT separates them into a gauge singlet and a triplet. 
Fundamentally the SMEFT describes a Higgs doublet in unbroken phase and its scalar manifold contains an $O(4)$ invariant fixed point~\cite{Alonso:2016oah} while the HEFT generally no longer requires this condition and can be reformulated into a SMEFT if and only there is an $O(4)$ invariant point. Thus
HEFT is more general than SMEFT.  
Diverse studies of HEFT appear recently \cite{Sun:2022snw,Sun:2022ssa,Graf:2022rco,Gomez-Ambrosio:2022why,Gomez-Ambrosio:2022qsi,Alonso:2021rac,Asiain:2021lch,Cohen:2021ucp,Herrero:2021iqt,Herrero:2022krh}.

The procedure that connect EFTs to UV-complete models is called \textsl{  matching}, which is an essential step for exploiting EFTs. For simplicity, we call matching HEFT to UV models as \textsl{HEFT matching}, similarly, matching SMEFT to UV models as \textsl{SMEFT matching}. 
Though SMEFT matching are well considered~\cite{Henning:2014wua, Drozd:2015kva, Chiang:2015ura, Huo:2015exa, Huo:2015nka, Brehmer:2015rna, Crivellin:2016ihg, Belusca-Maito:2016dqe, Dawson:2017vgm, Corbett:2017ieo, deBlas:2017xtg, Han:2017cfr, Jiang:2018pbd, Craig:2019wmo, Haisch:2020ahr, Gherardi:2020det, Dawson:2020oco, Marzocca:2020jze, Corbett:2021eux, Zhang:2021tsq, Zhang:2021jdf, Brivio:2021alv, Dedes:2021abc, Dawson:2021xei, Du:2022vso, Li:2022ipc, Dawson:2022cmu, Zhang:2022osj, Liao:2022cwh, 
Ellis:2023zim, Dawson:2023ebe, Li:2023cwy, Li:2023ohq, DasBakshi:2024krs, Dawson:2024ozw}~\footnote{Due to the rapid development of this area, we apologize for those work that we forget to cite here. Nowadays there are several tools developed for matching UV models to SMEFTs automatically: \textbf{MatchingTools}~\cite{Criado:2017khh} is a Python library for addressing EFT matching at tree level using functional
approach, \textbf{CoDEx}~\cite{DasBakshi:2018vni} and \textbf{Matchete}~\cite{Fuentes-Martin:2022jrf} are  Mathematica packages performing EFT matching upto 1-loop level utilizing functional methods, and \textbf{Matchmakereft}~\cite{Carmona:2021xtq} is a tool that automates the tree-level and one-loop matching via the plain Feynman diagram approach}, HEFT matching is less performed and only more commonly considered in composite Higgs models (CHMs)~\cite{Grojean:2013qca, Alonso:2014wta, Hierro:2015nna, Gavela:2016vte, Qi:2019ocx, Lindner:2022kxm}. A systematic study firstly appears in 
Ref.~\cite{Buchalla:2016bse}, which matches the SM model with an extra singlet to linear and non-linear EFTs and clarify their relations. A recent study of matching the SM with real singlet extension, complex singlet extension and 2 Higgs doublet model~(2HDM) to HEFT in  Ref.~\cite{Dawson:2023oce}, shows that different power countings (PCs) in HEFT can yield quite different results. Ref~\cite{Banta:2023prj} proposed to use ``straight-line'' basis of the 2HDM instead of \textit{Higgs basis} as in Refs~\cite{Dawson:2023ebe} and obtained a SMEFT-like EFT with better convergence property. Ref.s~\cite{Arco:2023sac, Buchalla:2023hqk} match 2HDM to EFT via diagrammatic method and functional method respectively and focus on the non-decoupling effects from the heavy states. In this work, we use functional method to match HEFT to general scalar extensions of the Standard Model.

The functional method starts from the Lagrangian of UV-complete models, then integrate out  heavy particles and get effective Lagrangian of light particles.
In SMEFT, the light Higgs together with Goldstone bosons appear in a doublet form. In HEFT, the light Higgs is a singlet together with a non-zero VEV, and Goldstone bosons are expresses in a unitary matrix as $U\equiv\exp\left(\frac{i \pi_i \sigma_i}{v}\right)$, where $\sigma_i$ are the three Pauli matrices, $\pi_i$ are massless Goldstone bosons.
%For SMEFT, in scalar sector it  Higgs doublet; for HEFT, it  Higgs singlet and a unitary matrix $U\equiv\exp\left(\frac{i \pi_i \sigma_i}{v}\right)$, where $\sigma_i$ are the three Pauli matrices, $\pi_i$
 %are massless Goldstone bosons.
 As we know Goldstone bosons always accompany with the non-zero VEV which breaks symmetry spontaneously, it is easy to be identified when there is only one non-zero VEV, just as in SM. In scalar extensions of the SM, there are generally two non-zero VEVs, Goldstone bosons are no longer associated with each VEV, but the combined electroweak VEV. We could identify them through the ``massless'' feature in linear form, however, since we would like to leave its exponential form while matching to HEFT, this method is infeasible.  
 How to identify Goldstone bosons in Lagrangian and later encapsulate them in a exponential form? This is a major obstacle when matching HEFT to general models having scalar extensions. 
 We solve this problem by considering $U$ matrix as a $SU(2)_L$ rotation and make 
rotated scalar representations in UV-complete model. 
 We also present how it is feasible and useful in real Higgs triplet model. Meanwhile it is easy to be applied for higher scalar multiplets such as quadruplet, septet, etc.  Through this representation matching HEFT to general scalar extensions become easy and trivial.

 The paper is organized as follows: we start by recapping the HEFT in section~\ref{sec:HEFT}. In section~\ref{sec:RHTE} based on real Higgs triplet model we firstly describe linear forms and its difficulty for HEFT matching, then we give its non-linear representation, i.e. $U$ representation and show a complete matching, finally we write $U$ representation in tensor notation. 
 In section~\ref{sec:general} we first give $U$ representations of complex triplet, quadruplets, then we write a general form for scalar multiplet extensions.
 Section~\ref{sec:con} is a short conclusion and discussion.

\section{HEFT\label{sec:HEFT}}
As discussed in the introduction, the HEFT treats the Higgs field and Golstone fields differently. Following the Callan-Coleman-Wess-Zumino (CCWZ) construction~\cite{Coleman:1969sm}, the EW would-be Goldstones $\pi_i$ are embeded into a dynamical unitray matrix $U\equiv\exp\left(\frac{i \pi_i \sigma_i}{v}\right)$, where $\sigma_i$ are the three Pauli matrices. Inspired by the similarity between the EW Goldstone bosons and pions, the HEFT operators can be constructed order by order in the number of the covariant derivatives (therefore the HEFT is also known as the electroweak chiral Lagrangian with a light Higgs boson). At the lowest order, the HEFT Lagrangian is given as,
\begin{align}
\mathcal{L}_{\text{HEFT}}^{\text{LO}}\supset&\frac{1}{2}\mathcal{K}(H_0)\partial_\mu H_0 \partial^\mu H_0-\mathcal{V}(H_0)+\frac{\vev^2}{4} \mathcal{F}(H_0)\langle D_\mu U^\dagger D^\mu U\rangle \nonumber \\
&+\frac{\vev^2}{4}\mathcal{G}(H_0)\langle U^\dagger D_\mu U\sigma_3 \rangle \langle U^\dagger D^\mu U\sigma_3 \rangle \nonumber \\
&-\frac{\vev}{\sqrt{2}}(\bar{Q}_L U\mathcal{Y}_Q(H_0) Q_R)+\bar{L}_L U\mathcal{Y}_L(H_0)L_R + h.c.)~,
\label{eq:HEFT}
\end{align}
where $\vev=246$~GeV represents the vacuum expectation value (VEV) of the Higgs field, $D_\mu$ is the covariant derivative, and $\mathcal{K}(H_0)$, $\mathcal{V}(H_0)$, $\mathcal{F}(H_0)$, $\mathcal{G}(H_0)$, and $\mathcal{Y}_{Q/L}(H_0)$ are polynomial functions of $H_0$ with the following forms,
\begin{gather}
    \mathcal{K}(H_0)=1+c_1^k\frac{H_0}{\vev}+c_2^k\frac{H_0^2}{\vev^2}+\cdots, \\
    \mathcal{V}(H_0)=\frac{1}{2}m_{H_0}^2 H_0^2\left[1+(1+\Delta\kappa_3)\frac{H_0}{\vev}+\frac{1}{4}(1+\Delta\kappa_4)\frac{H_0^2}{\vev^2}+\cdots\right],  \\
    \mathcal{F}(H_0)=1+2(1+\Delta a)\frac{H_0}{\vev}+(1+\Delta b)\frac{H_0^2}{\vev^2}+\cdots, \\
    \mathcal{G}(H_0)=\Delta\alpha+\Delta a^\slashed{C}\frac{H_0}{\vev}+\Delta b^\slashed{C}\frac{H_0^2}{\vev^2}+\cdots, \\
    \mathcal{Y}_Q(H_0)=\text{diag}\left(\sum_n Y_U^{(n)}\frac{H_0^n}{\vev^n},\,\sum_n Y_D^{(n)}\frac{H_0^n}{\vev^n}\right),\ \mathcal{Y}_L(H_0)=\text{diag}\left(0,\,\sum_n Y_\ell^{(n)}\frac{H_0^n}{\vev^n}\right).
\end{gather}
The normalization function $\mathcal{K}(H_0)$ of the Higgs kinetic term generally is redundant. One can perform a field redefinition $H'_0\rightarrow\int_0^{H_0}\sqrt{\mathcal{K}(s)}ds$, and reabsorb its entire effects into redefinitions of the other functions. But we observe that such a form of the Higgs kinetic normalization is commonly generated in the middle step of a UV and EFT matching process. Therefore we present it without performing the redefinition which is usually complicated. The second term contains the Higgs potential; the third term describes the kinetic terms of the gauge bosons; the fifth terms accounts for the Yukawa interactions in which the $n=0$ terms yield fermion masses. In this work we will focus on the bosonic sector, therefore we will not consider the $\mathcal{Y}_{Q, L}$ terms anymore, which does not impact our results. 

Now let us comment on the fourth term, which violates the custodial symmetry. It is usually considered as a next-to-leading order (NLO) term (a chiral dimension 4 term) since it is only generated at the loop level in the SM. However, this is an assumption on the UV theory and breaks the power-counting. It is definitely possible that, though from the experiments we know that the custodial symmetry violation is rather small, the custodial symmetry breaking can be triggered by new physics at the electroweak scale.  Thus from the point of view of effective field theory, this operator is characterized by the same scale as the others and should be kept at the leading order (LO) without introducing any UV assumptions. 

\section{The SM with one real Higgs triplet extension (RHTE)~\label{sec:RHTE}}

In the RHTE besides the $SU(2)_L$ doublet there exists a triplet with hyper-charge $Y=0$. 
Since singlet and doublet extension won't break custodial symmetry, it is the minimum extension of custodial-symmetry-violation model. 
%Even though, 
%the procedure of matching HEFT to scalar extensions is similar and we get a way 
%which is easily generalized to higher multiplet extensions.
In this section, firstly we introduce linear forms of the Lagrangian, which  
describe same UV model but have distinctions for matching.  
%but has significant distinction for matching.
%Then we rotate them by $U$ matrix in $U$ representation
%and show the whole procedure of matching, 
Then we give a non-linear form, $U$ representation, which separate out $U$ matrix  explicitly and thus is feasible
 for HEFT matching. We further show a complete matching procedure. 
Finally we introduce the tensor notation and generalize $U$ representation to general scalar multiplet extensions. 

\iffalse
Usually it is written in a matrix as $\Sigma= \frac{1}{2} \Sigma_i \sigma_i$,$i=1,2,3$, $\sigma_i$ are Pauli matrices and $\Sigma_i$ are three scalar fields. 

Due to $\sigma_i$ matrices are not only generators of $SU(2)_L$ under doublet representation
but also constitute a triplet representation, known as associated representation, the $\Sigma$ form  makes Lagrangian compact. However, this form is only suitable for triplets, tensor form~(cite quadrplet, septet tensor representation papers) are more general for multiple scalar representations. So we firstly  match RHTE to HEFT in $\Sigma$ form, 
then we show how the same idea is suitable in tensor form. In next subsections, we first introduce the RHTE in linear form, both doublet and bidoublet form are presented, then we discuss rotated representations. Finally we ...

\fi

%\subsection{$\sigma$-basis}
\subsection{Linear forms}
\subsubsection{Doublet}
Usually the Lagrangian of RHTE in scalar sector can be expressed as~\cite{Corbett:2021eux,Ellis:2023zim}
\begin{align}
{\cal L}_{\rm RHTE}(\mathrm{H},\Sigma)\supset \left(D_{\mu}{\rm H} \right)^{\dagger}\left(D^\mu {\rm H}\right)+\langle D_{\mu}\Sigma^{\dagger} D^{\mu}\Sigma\rangle -V\left( H,\Sigma \right),
\label{LHSigma}
\end{align}
with 
\bea
V\left({\rm H},\Sigma \right)&=&
Y_1^2  \mathrm{H}^\dagger \mathrm{H} +Z_1  (\mathrm{H}^\dagger \mathrm{H})^2
+ Y_2^2 \langle \Sigma^\dagger \Sigma \rangle
+  Z_2 \langle \Sigma^\dagger \Sigma \rangle^2 \nonumber \\
&&+ Z_3 \mathrm{H}^\dagger \mathrm{H} \langle \Sigma^\dagger\Sigma \rangle
+2 Y_3  \mathrm{H}^\dagger \Sigma\mathrm{H}~,
\label{Hpotential}
\eea
where $\langle ... \rangle$ denotes the trace, 
$Y_is,i=1,2,3$ are dimensional parameters while $Z_is,i=1,2,3$ are dimensionless. ${\rm H}$ is  
the SM Higgs doublet and $\Sigma$ is the real triplet, which could be expressed as 
\begin{align}
{ \rm H}=
\begin{pmatrix}
G^+\\
\frac{1}{\sqrt{2}}\left(v_{\rm H}+h+iG^0\right)
\end{pmatrix},\hspace{1cm}
\Sigma=\frac{1}{2} \Sigma_i \sigma_i
=
\frac{1}{2} 
\begin{pmatrix}
v_\Sigma+\Sigma^0  & \sqrt{2}\Sigma^+\\
\sqrt{2}\Sigma^- & - v_\Sigma-\Sigma^0 
\end{pmatrix}, i=1,2,3,
\label{Sigmamatrix}
\end{align}
where
$v_{\rm H}$, $v_\Sigma$ are vacuum expectation values (VEVs) after spontaneously symmetry breaking (SSB). 
$G^0$ is a pseudoscalar which does not mix with the scalar $h$ except for $CP$ violation. 
$G^+$ is a charged scalar. 
 $\Sigma_i$s constitute 3 triplet components, $\sigma_i$s are Pauli matrices. 
 $\Sigma^0= -v_\Sigma+\Sigma_3 $ is the shifted field along VEV,   $\Sigma^{\pm}=\frac{1}{\sqrt{2}}(\Sigma_1\mp i \Sigma_2)$ are canonically normalized charged scalars.

Due to two VEVs, $G^\pm$ and $\Sigma^\pm$ should be mixed and give massless Goldstones and massive charged Higgs, which are 
\begin{align}
\left(\begin{array}{c}
 G_{\rm EW} ^+ \\
 H^+\\
\end{array} \right)
= 
\left(\begin{array}{cc}
 \cos \delta & -\sin \delta \\
 \sin \delta & \cos \delta 
\end{array} \right)
\left(\begin{array}{c }
 G^+ \\
 \phi^+\\
\end{array} \right) ,  
\label{Gphimix}
\end{align}
where $G_{\rm EW} ^+$ represents Goldstones perpendicular to electroweak vacuum, the tangent of rotation angle is 
\begin{align}
    \tan \delta =  2 v_\Sigma /v_{\rm H}.
    \label{tandelta}
\end{align}

Under $SU(2)_L\times U(1)_Y$ gauge symmetry, the doublet ${\rm H}$ and the real triplet $\Sigma$ transform as 
\begin{align}
   {\rm H}\to \mathfrak{g}_L(\boldsymbol{\theta}) \exp(\frac{i}{2} \theta_Y)~{\rm H} ~ ,~\quad \Sigma\to \mathfrak{g}_L(\boldsymbol{\theta}) 
    ~\Sigma ~ \mathfrak{g}_L(\boldsymbol{\theta})^\dagger~,  \quad \mathfrak{g}_L(\boldsymbol{\theta})= { \rm exp}(\frac{i}{2} \theta_i \sigma_i),
\end{align}
which explain why the potential has the trilinear interaction $\mathrm{H}^\dagger \Sigma\mathrm{H}$ in Eq.~\eqref{Hpotential} . 
\subsubsection{Bidoublet}
Not only the triplet could be given in Pauli matrices, the complex doublet could also be given as a matrix, which is denoted as bidoublet and formally it is
\begin{align}
     S \equiv (\tilde{\rm H}, {\rm H}) = \left( \begin{array}{cc}
    \frac{1}{\sqrt{2}}(v_{\rm H}+h-i G^0) & G^+\\
    -G^-  & \frac{1}{\sqrt{2}}(v_{\rm H}+h +i G^0)
    \end{array}
    \right)
    = \frac{1}{\sqrt{2}}\left(
    (v_{\rm H}+h) \mathbb{I} + i G_i \sigma_i\right )~,
    \label{Slinearform}
\end{align}
where in the last term we rewrite the bidoublet $S$ in Pauli matrices plus an identity matrix. 
$G^\pm=\pm \frac{i}{\sqrt{2}} (G_1 \mp i G_2)$, which is not equal to $\frac{1}{\sqrt{2}} (G_1 \mp i G_2)$ just as $\Sigma^\pm$ are, but has one more $i$ factor. 
It is because 
 the bidoublet form exerts this extra $i$. Anyhow, $G^\pm$ could mix with $\Sigma^\pm$ so that $G_i$s has a crossed correspondence to $\Sigma_i$s, e.g., $G_1 \sim \Sigma_2$ and $G_2 \sim \Sigma_1$.

Under $SU(2)_L\times U(1)_Y$ symmetry, the bidoublet $S$  transforms as \begin{align}
   S\to \mathfrak{g}_L(\boldsymbol{\theta}) ~S ~ \mathfrak{g}_Y(\theta_Y)^\dagger,~\quad  
   \mathfrak{g}_L(\boldsymbol{\theta})= \rm{exp}(\frac{i}{2} \theta_i \sigma_i), \quad
   \mathfrak{g}_Y(\theta_Y)=
   \rm{exp}(\frac{i}{2}
   \theta_Y \sigma_3).
\end{align}

The Lagrangian of the RHTE is 

\begin{align}
{\cal L}_{\rm RHTE}( S ,\Sigma)\supset
\langle D^\mu  S^\dagger D_\mu  S \rangle~
+\langle D_{\mu}\Sigma^{\dagger} D^{\mu}\Sigma\rangle-V(  S ,\Sigma ),
\end{align}
\begin{align}
   V(  S ,\Sigma )
=  Y_1^2\langle  S ^\dagger  S  \rangle   +
Z_1 \langle  S ^\dagger  S  \rangle^2 +
Y_2^2 \langle \Sigma^\dagger \Sigma \rangle +
Z_2 \langle \Sigma^\dagger \Sigma \rangle^2 + 
Z_3 \langle  S ^\dagger  S  \rangle \langle \Sigma^\dagger \Sigma \rangle - Y_3 \langle
S^\dagger \Sigma  S  \sigma_3 \rangle~.
\end{align}
Where  $Y_i$s and $Z_i$s are same as in Eq.~\eqref{Hpotential}.

The Lagrangian ${\cal L}_{\rm RHTE}( S ,\Sigma)$ is equivalent to ${\cal L}_{\rm RHTE}(\mathrm{H},\Sigma)$ as in Eq.~\eqref{LHSigma}. When matching the RHTE to SMEFT, it mostly use ${\cal L}_{\rm RHTE}(\mathrm{H},\Sigma)$ but not ${\cal L}_{\rm RHTE}(S,\Sigma)$, because SMEFT is usually expressed in $\mathrm{H}$. Principally if start with ${\cal L}_{\rm RHTE}(S,\Sigma)$ we also get SMEFT since the conversion between $\mathrm{H}$ ans $S$ is equivalent for physics and choose which is just a convention.
%However, matching to SMEFT is still feasible principally if start with ${\cal L}_{\rm RHTE}(S,\Sigma)$. 
%Just need one more step that replacing $S$ term with equivalent ${\rm H}$ term in the last. 
In both Lagrangians $\Sigma$ are linearly connected to $S$ or ${\rm H}$, the procedure of integrating out it is similar.  
 The situation is different for HEFT matching. 
 %Through ${\cal L}_{\rm RHTE}(S,\Sigma)$ 
 In Eq.~\eqref{Gphimix}
 we see Goldstones are mixing states of $G^\pm$ and $\Sigma^\pm$, integrating out $\Sigma$ means there leaves fake Goldstone bosons in EFT. 
 There could be an approximation, e.g., we firstly get SMEFT and then write doublet ${\rm H}$ in polar coordinates to get HEFT. Meanwhile, through this way we could constrain HEFT through SMEFT results~\cite{Anisha:2024xxc}. 
Nevertheless, this procedure results in a constrained HEFT, not a general one. 
 Generally we should pick out real Goldstone bosons and make them in exponential form, both ${\cal L}_{\rm RHTE}(\mathrm{H},\Sigma)$ and ${\cal L}_{\rm RHTE}(\mathrm{H},\Sigma)$ are not suitable. We would better get a new form for HEFT matching.

\subsection{$U$ representation}
In linear forms, the Goldstone bosons, light and heavy scalars are all expressed linearly, it is difficult to encapsulate Goldstone bosons into exponential form. In this section we introduce a $U$ representation, which separate Goldstone bosons from other scalars at the beginning so that HEFT matching could be carried on.

\subsubsection{The model}
In UV-complete model, there are three Goldtone bosons after SSB of $SU(2)_L\times U(1)_Y$. 
They are expressed 
in exponential form as $U\equiv\exp\left(\frac{i \pi_i \sigma_i}{v}\right)$ in HEFT. 
For a straightforward HEFT matching that just integrate out heavy states, it is necessary to 
make $U$ matrix explicit in Lagrangian.
%To leave them in exponential form as $U\equiv\exp\left(\frac{i \pi_i \sigma_i}{v}\right)$ 
%after integrating the heavy states, 
%we pick them out in the UV-complete model before matching.

Since $U$ matrix could also be considered as an $SU(2)_L$ rotation of angle $\boldsymbol{\pi}/v$
, we redefine the bidoublet and the triplet as 
\bea
    S_R&\equiv& U R,\qquad R= \frac{1}{\sqrt{2}}\left((v_H+h) \mathbb{I}+ i \hat{\rho}\right), \quad \hat{\rho}=  \rho_i \sigma_i
    ~,\label{SRmatrix}\\
    \Sigma_\Phi &\equiv&
U \Phi U^{\dagger},\quad\Phi=\frac{1}{2}\phi_i \sigma_i~
=
\frac{1}{2} 
\begin{pmatrix}
v_\Sigma+\phi^0  & \sqrt{2}\phi^+\\
\sqrt{2}\phi^- & - v_\Sigma-\phi^0 
\end{pmatrix},
\label{Phimatrix}
\eea
where $v_{\rm H}$ and $v_\Sigma$ are VEVs of the doublet and triplet, $R$ and $\Phi$ keep the linear forms of bidoublet and triplet respectively (see Eq.s~\eqref{Slinearform}\eqref{Sigmamatrix}), $\rho^\pm=\pm \frac{i}{\sqrt{2}}(\rho_1\mp i \rho_2) $ are heavy charged Higgs, $\phi^0 = -v_\Sigma + \phi_3$, $\phi^{\pm}=\frac{1}{\sqrt{2}}(\phi_1\mp i \phi_2)$ are also charged Higgs. Since $U$ represent massless Goldstones,
$\rho^\pm$ should be proportional to $\phi^{\pm}$ so that the number of free states is kept to be seven as in liner form. The proportionality factor could be settled down by cancelling kinetic mixing between Goldstones and charged Higgs.

%$S_R$ and $\Sigma_\Phi$ are  bidoublet and triplet under $SU(2)_L$ gauge symmetry, 
%$R$ is a $2\times 2$ matrix which keeps the linear form of bidoublet as in Eq.~\eqref{Slinearform}, $\Phi$ also keeps the linear form of triplet (see Eq.~\eqref{Sigmamatrix}).
%However, $R$ and $\Phi$ are not bidoublet and triplet under $SU(2)_L$ gauge transformation, which we give out details later.  $\rho_i$ in $R$ are heavy states, which have linear relations with $\phi_i$s. $\phi^0 = -v_\Sigma + \phi_3$, $\phi^{\pm}=\frac{1}{\sqrt{2}}(\phi_1\mp i \phi_2)$.

According to linear form,
the Lagrangian is  
\begin{align}
{\cal L}_{\rm RHTE}( S_R ,\Sigma_\Phi)\supset
\langle D^\mu  S_R^\dagger D_\mu  S_R \rangle~
+\langle D_{\mu}\Sigma_\Phi^{\dagger} D^{\mu}\Sigma_\Phi\rangle-V( S_R ,\Sigma_\Phi ),
\end{align}
\begin{align}
   V(  S_R ,\Sigma_\Phi )= V(  R ,\Phi )
&=  Y_1^2\langle  R ^\dagger  R  \rangle   +
Z_1 \langle  R ^\dagger  R  \rangle^2 +
Y_2^2 \langle \Phi^\dagger \Phi \rangle +
Z_2 \langle \Phi^\dagger \Phi \rangle^2 \nonumber  \\
&+ 
Z_3 \langle  R ^\dagger  R  \rangle \langle \Phi^\dagger \Phi \rangle - Y_3 \langle
R^\dagger \Phi  R  \sigma_3 \rangle~,
\label{RHTEPotential}
\end{align}
where $V(S_R ,\Sigma_\Phi)$ does not depends on $U$ matrix at all, so it degenerates to $V(R ,\Phi )$. But $U$ matrix 
moves into kinetic terms. We get simpler potential in expense of more complex kinematics. For example, the kinetic term of $S_R$ have many $U$ matrices,

%Potential becomes simpler in expense of complex kinematics, e.g., the kinetic term of $S_R$ have many $U$ matrices. 
\bea
    \mathcal{L}^{\text{kin}}_{S_R}&=&
\frac{1}{2}\langle D^\mu S^\dagger D_\mu S\rangle~ \nonumber \\
&=& 
 \frac{1}{2}( \langle D_\mu R^\dagger D^\mu R \rangle
  +
  \langle (RR^\dagger)D^\mu U^\dagger D_\mu U \rangle 
  + \langle   U^\dagger D_\mu U (R D^\mu R^\dagger -D^\mu R  R^\dagger)\rangle) \\
&=& -\frac{i}{4} v_{\rm H}
 \langle  U^\dagger D^\mu U (D_\mu \hat{\rho}+ D_\mu \hat{\rho}^\dagger) \rangle~+...,
\label{Phimix1}
\eea
where in the last line we express the kinetic mixing term explicitly and leave others in suspension points. It gives a kinetic mixing between $\pi_i$s and $\rho_i$s, which is proportional to
the VEV $v_{\rm H}$. This kinetic mixing could cause $\pi_i$s no longer massless and thus not real Goldstone bosons. It should be removed, otherwise this representation is not suitable for a simple HEFT matching. 
%since we would like to take $U$ matrix as real Goldstones.
%It seems to be an dangerous thing that prevent $U$ to be real Goldstone bosons.

The mixing term from $\Sigma_\Phi$ is 
\begin{align}
    \mathcal{L}^\text{mixing}_{\Sigma_\Phi}&=
   \frac{1}{2}v_\Sigma \langle  U^\dagger D^\mu U  ([\sigma_3, D_\mu \Phi] +[\sigma_3, D_\mu \Phi^\dagger]) \rangle~.
   \label{Sigmamix1}
   \end{align}
Comparing Eq.s\eqref{Phimix1}\eqref{Sigmamix1}, we get 
\begin{align}
    \hat{\rho}=-2i\frac{v_\Sigma}{v_{\rm H}}[\sigma_3, \Phi]=
    -2\frac{v_\Sigma}{v_{\rm H}}( \phi_2 \sigma_1 -\phi_1 \sigma_2 )
    \label{hatrho}
\end{align}
where $[\sigma_3,D_\mu \Phi]=D_\mu[\sigma_3, \Phi]$ has been used due to $\Phi$ transforms  as in Eq.~\eqref{PhiTransformation} and thus its $D_\mu$ operator commutes with $\sigma_3$. In this way we eliminate the kinetic mixing between $\pi_i$s and heavy states, and get $\rho_i$s. 
From Eq.~\eqref{hatrho}, we also see 
\begin{align}
\rho^\pm = 2\frac{v_\Sigma}{v_{\rm H}}  \phi^\pm ,\quad  \rho_3 =0 
\label{rhotriplet}
\end{align}
which means we actually put the charged Higgs in the bidoublet matrix by a proportional factor of $ 2 v_\Sigma/{v_{\rm H}}$, and the neutral part is zero because neutral part of the real triplet has no effect to Goldstones. It is interesting to see $ 2 v_\Sigma/{v_{\rm H}}$ is equal to  $\tan \delta$ in Eq.~\eqref{tandelta}, which represents the rotation angle between $G^+$ and $\Sigma^+$. 
Apparently the rotation between charged states does not vanish but hide in new form of doublet.

%In other words, 
%in linear form the mixing between charged states in potential but in $U$ representation it appear in kinetic part.  

\iffalse
Explicitly $\rho_i$s are
\begin{align}
     \rho_1= -2 \frac{v_\Sigma}{v_{\rm H}} \phi_2, \quad 
     \rho_2= 2 \frac{v_\Sigma}{v_{\rm H}} \phi_1,
     \quad 
     \rho_3=0~.
     \label{rho1}
 \end{align}
 \fi

%\subsubsection{The gauge transformations}
\iffalse
In $U$ representation
$R$ and $\Phi$ are written explicitly as 
 
 \begin{align}
R= \frac{1}{\sqrt{2}}((v_{\rm H}+h) \mathbb{I}-2 i \frac{v_\Sigma}{v_{\rm H}} (\phi_2 \sigma_1- \phi_1 \sigma_2))~,~ \Phi = \frac{1}{2} \phi_i \sigma_i  
\label{Rreal}
\end{align}
\fi

At last let us see how $R$ and $\Phi$ transforms under gauge symmetry. We start from 
the gauge transformations of $S_R$ and $\Sigma_\Phi$, which are
\begin{align}
   S_R\to \mathfrak{g}_L(\boldsymbol{\theta}) ~S_R ~ \mathfrak{g}_Y(\theta_Y)^\dagger, \quad \Sigma_\Phi\to  \mathfrak{g}_L(\boldsymbol{\theta}) ~\Sigma_\Phi ~ \mathfrak{g}_L(\boldsymbol{\theta})^\dagger.
\end{align}
As we know $U$ transforms as~\cite{Sun:2022ssa}
\bea
   U\to \mathfrak{g}_L(\boldsymbol{\theta}) 
   ~U~ \mathfrak{g}_Y(\theta_Y)^\dagger, \label{Utransformation}
\eea
so that $R$ and $\Phi$ should transform as 
\bea
   R \to \mathfrak{g}_Y(\theta_Y) ~R~ \mathfrak{g}_Y(\theta_Y)^\dagger\label{Rtransformation}\label{RTransformation},\quad
   \Phi\to \mathfrak{g}_Y(\theta_Y) 
    ~\Phi ~ \mathfrak{g}_Y(\theta_Y)^\dagger.
\label{PhiTransformation}
\eea
Which shows $R$ and $\Phi$ coincidentally share same gauge transformation. 
This is satisfied while $R$ actually is
\begin{align}
R= \frac{1}{\sqrt{2}}((v_{\rm H}+h) \mathbb{I}-2 i \frac{v_\Sigma}{v_{\rm H}} 
[\sigma_3, \Phi]. 
\end{align}

\subsubsection{Matching}
We simply express the convenience of $U$ representation in the matching process.
In RHTE generally we have 6 free parameters, among them we choose a parameter set
\begin{align}
    \left(Z1,Z2,Z3,Y3,v_H,v_\Sigma \right)
\end{align}
for study as it is convenient and easy to be calculated. 
Constraints such as mass of the light SM-like Higgs is equivalent to 125~GeV may possibly cause some inconsistency, however, we leave these subtleties for further studies.
We choose 
\begin{align}
    \xi\equiv v_\Sigma/v_H
\end{align} 
for power counting as it is a small quantity less than 0.02~\cite{Cheng:2022hbo} by experimental constraints. It also indicates that custodial symmetry violation is small. 
There are 4 fields left so far: $h$, $\phi^0$, $\phi^+$,$\phi^-$. Among them, $h$ mixes with $\phi^0$ and gives a light Higgs $H_0$ and heavy neutral Higgs $K_0$, which is 
\beq
\left(\begin{array}{c}
 H_0 \\
 K_0 \\
\end{array} \right)
= 
\left(\begin{array}{cc}
 \cos \gamma & -\sin \gamma \\
 \sin \gamma & \cos \gamma 
\end{array} \right)
\left(\begin{array}{c }
 h^0 \\
 \phi^0\\
\end{array} \right),
\label{H_0K0Rotation}
\eeq
where $\sin \gamma= 2 \xi + \frac{2 (4 Z_1 -Z_3)\xi^2 v}{Y_3} 
+\mathcal{O}(\xi^3)~$. 

We get equation of motions (EoMs) for the heavy Higgs and solve them in series of $\xi$. Specifically, 
the heavy Higgs states are thus expanded as\footnote{A general solution should be a Laurent series that include negative exponents. But as a quantum field we do not want a divergent field configuration around $\xi=0$.} 
\bea
K_0  &= & K_{0 0} + \xi K_{0 1} + \xi^2 K_{0 2}+... \nonumber \\
H_1 &=& H_{10} + \xi H_{11} + \xi^2 H_{12}+... \nonumber \\
H_2 &=& H_{20} + \xi H_{21} + \xi^2 H_{22}+...~,
\label{expandxi}
\eea
and solved order by order. Under $U$ representation, the $U$ and heavy Higgs are separate already, so that the solution is obtained as simple as in polynomial functions. 
Finally the HEFT are obtained as 

\bea
\mathcal{L}_{\rm HEFT}&\subset&
\frac{1}{2}(1+ \frac{2\xi^2}{ v_{\rm H }^2} H_0^2)D_\mu H_0 D^\mu H_0 -\mathcal{V}(H_0)\\ &+& \frac{(v_{\rm H}+H_0)^2}{4}\left(1+\frac{4\xi^2}{ v_{\rm H }^2}(v_{\rm H}^2+ v_{\rm H} H_0  + H_0^2) \right) \langle D_\mu U^\dagger D^\mu U\rangle \\
&+& \frac{\xi^2 (v_{\rm H}+H_0 )^4}{2 v_{\rm H }^2}  \langle  U^{\dagger} D_\mu U
\sigma_3 \rangle  \langle U^{\dagger} D^\mu U \sigma_3\rangle  +\mathcal{O}(\xi^3)+..., 
\eea
 which corresponds to the standard HEFT LO form (Eq.~\eqref{eq:HEFT}) with the coefficients listed in Tab.~\ref{tab:HEFT_couplings}. $\Delta\kappa_3$ and $\Delta\kappa_4$ are obtained from $\mathcal{V}(H_0)$, which we do not give an explicit expression due to its long polynomials. At $\mathcal{O}(\xi^3)$ oder, there appears operators with 4 derivatives such as 
  $\langle D_\mu U^\dagger D^\mu U\rangle\langle D_\nu U^\dagger D^\nu U\rangle$,
  $\langle  U^{\dagger} D_\mu U
\sigma_3 \rangle  \langle U^{\dagger} D^\mu U \sigma_3\rangle \langle D_\nu U^\dagger D^\nu U\rangle$, $\langle  U^{\dagger} D_\mu U
\sigma_3 \rangle  \langle U^{\dagger} D_\nu U \sigma_3\rangle \langle D^\mu U^\dagger D^\nu U\rangle$, etc. Since we focus on introduction of $U$ presentation in this paper, we leave the detailed results for further study. 
\begin{table}[h!]
\centering
\begin{tabular}{ccccccccc}
\hline\hline
$\Delta a$ & $\Delta b$ & 
$\Delta\alpha$ & $\Delta a^\slashed{C}$ & 
$\Delta b^\slashed{C}$ & $c_1^k$ & $c_2^k$ &$\Delta\kappa_3$ & $\Delta\kappa_4$ 
\\
\hline\hline
$4\xi^2$ & $16\xi^2$  &
$2\xi^2$ & $8\xi^2$ & $12\xi^2$ & $0$ & $2\xi^2$ & $-2\left(2-\frac{Z_3}{Z_1}\right)\xi^2$ & $-12\left(2-\frac{Z_3}{Z_1}\right)\xi^2$ 
\\
\end{tabular}
\caption{HEFT couplings for the triplet model. All the couplings are shown up to $\mathcal{O}(\xi^2)$.~\label{tab:HEFT_couplings}}
\end{table} 

\subsection{Tensor notation and $U$ representation}
In last section we have shown 
in $U$ representation  matching HEFT to RHTE could be carried on smoothly. 
However, there leaves a drawback that the triplet $\Sigma\propto \Sigma_i \sigma_i$ form is special for triplet and not general for other scalar multiplets, e.g. quadruplet. Fortunately,
this could be made up by using tensor notation instead, because tensor notation is general for all scalar multiplets. Furthermore,
the idea of rotating 
heavy states with $U$ matrix could still be applied in tensor notation, which is illustrated detailedly in this section.
 
\iffalse
Generally the scalar multiplets are written in tensor notation, such as quadruplet in Ref.~cite~1705.0255,1704.07851,1711.10391,2311.17995~, and septet in Ref.~cite(1301.6455). The $\Sigma$
 form (
$\Sigma=1/2 \Sigma_i \sigma_i$, $i=1,2,3$) for triplet is an special example due to $\sigma_i$'s property. 
Although $U$ presentation is successful in $\Sigma$ form, we should prove it is also applicable in tensor notation, so that it could be used for general scalar extensions. 
\fi
In tensor notation, a triplet is given by a symmetric tensor $\Sigma_{ij}=\Sigma_{(ij)}$, where $i(j)=1,2$ represent two components of a doublet representation. $\Sigma_{ij}$ constitute three components in highest weight representation, which correspond to
 $\Sigma_{11}= -\Sigma^+$, 
$\Sigma_{12}= \Sigma_3/\sqrt{2}=(v_\Sigma+ \Sigma^0)/\sqrt{2}$, 
$\Sigma_{22}= \Sigma^-$ as in Eq.~\eqref{Sigmamatrix}.
%The idea of $U$ representation is expressing $U$ explicitly in the field states.
To separate out $U$ matrix, we decompose $\Sigma_{ij}$ into $U$ matrix and $\phi_{ij}$
according to its gauge transformation under $SU(2)_L$, which is  
%which use similar ideas as in \eqref{Phimatrix} but have different form.  
\beq
\Sigma_{ij} = U_i^k U_j^l \phi_{kl}, 
\eeq
where $\phi_{kl}$s represent three heavy scalar states. They constitute a triplet tensor under $SU(2)$ symmetry, but only gauged by $U(1)_Y$ similarly as in Eq.~\eqref{PhiTransformation}. 

Since tensor is principally symmetric products of doublets, for the doublet scalar doublet form is more convenient than bidoublet form in this situation. 
%Instead of bidoublet, we find that doublet ${\rm H}$ is more suitable for tensor notation, since tensor is principally made of doublets. 
Similarly we decompose the doublet as 
\beq
{\rm H}_i = U_i^j   \mathfrak{h}_j,\quad  \mathfrak{h}= \left ( \begin{array}{c}
  \chi^+\\
  \frac{1}{\sqrt{2}}\left(v_{\rm H}+h+i\chi^0\right)\\
\end{array} 
\right ), 
\label{RotatedDoublet}
\eeq
where $\chi^\pm,\chi^0$ are  also heavy scalar states. 
Due to number of free scalars are seven in RHTE, $\chi^\pm,\chi^0$ must be some linear combinations of  
$\phi_{ij}$s,
which
should be determined by kinetic mixing criteria.
It is easy to guess that $\chi$ fields are proportional to $v_\Sigma$ (see Eq.~\eqref{hatrho}), which means $\chi$ fields become zero if $v_\Sigma=0$, and ${\rm H}_i$ returns to SM doublet. Even though we do not use bidoublet form, $\tilde{\rm H}_i$ is also important for tensor notation. Here we get 
$\tilde{\rm H}= i \sigma_2 U^\ast \mathfrak{h}^\ast =
U \tilde{\mathfrak{h}}$.

The Lagrangian is
\iffalse
\begin{align}
{\cal L}_{\rm RHTE}( {\rm H}_i ,\Sigma_{ij})\supset
 D^\mu {\rm H}^\dagger D_\mu  {\rm H}
+\frac{1}{2}\langle D_{\mu}\Sigma^{\dagger} D^{\mu}\Sigma \rangle-V({\rm H} ,\Sigma ),
\end{align}
\fi

\begin{align}
{\cal L}_{\rm RHTE}( {\rm H}_i ,\Sigma_{ij})\supset
 D^\mu {\rm H}^{\ast i} D_\mu  {\rm H}_i
+\frac{1}{2} D_{\mu}\Sigma^{\ast ij} D^{\mu}\Sigma_{ij} -V({\rm H} ,\Sigma ),
\end{align}

where $\Sigma$ represents $\Sigma_{ij}$,
%similarly $\Phi$ represents $\phi_{ij}$, 
the potential is
\begin{align}
   V( {\rm H} ,\Sigma  )
=  Y_1^2 \mathfrak{h}^{\ast i} \mathfrak{h}_i +
Z_1  (\mathfrak{h} ^{\ast i}\mathfrak{h}_i)^2 +
Y_2^2  \phi^{\ast ij } \phi_{ij}  +
Z_2 (\phi^{\ast ij } \phi_{ij})^2 + 
Z_3 \mathfrak{h}^{\ast i} \mathfrak{h}_i  \phi^{\ast jk } \phi_{jk}  +2 Y_3\left(  \phi^{\ast i j} \tilde{\mathfrak{h}}_i \mathfrak{h}_j  + h.c. \right),
\end{align}
where %$\langle \phi^\dagger \tilde{\mathfrak{h}} \mathfrak{h}^T \rangle= \phi^\ast_{ij} \tilde{\mathfrak{h}}^i \mathfrak{h}^j$, 
no $U$ matrix appears in potential because they are canceled in each term.  
In each kinetic term there exist kinetic mixing, next goal is to make two kinetic mixings cancel so that we get a framework easy to calculate.  
The mixing terms are

\bea
\mathcal{L}^\text{mix}_{\rm H}&=& 
 \langle  \mathfrak{h}_2 \rangle \Bigl( (U^\dagger D_\mu U)_1^{2} D^\mu \mathfrak{h}^{\ast 1} - (U^\dagger D_\mu U)_2^{1} D^\mu \mathfrak{h}_{1} %\nonumber \\ & &
 +
(U^\dagger D_\mu U)_2^{2} (D^\mu \mathfrak{h}^{\ast 2}-
D^\mu \mathfrak{h}_{ 2})
\Bigr),
\label{Hmix3}
\eea
 
\bea
\mathcal{L}^\text{mix}_{\Sigma}&=&
 \langle \phi_{12} \rangle   \Bigl (  (U^\dagger D_\mu U)_i^1  D^\mu \phi^{\ast i 2 } + (U^\dagger D_\mu U)_i^2  D^\mu \phi^{\ast i 1 } \nonumber \\
 &&\qquad\quad  - 
 (U^\dagger D_\mu U)_1^i  D^\mu \phi_{ i 2 } - (U^\dagger D_\mu U)_2^i  D^\mu \phi_{ i 1 }  \Bigr ) ,
\label{RTmix}
\eea
where $\langle  \mathfrak{h}_2 \rangle = \langle  \mathfrak{h}^{\ast 2} \rangle$, $\langle  \phi_{12} \rangle = \langle  \phi^{\ast 12} \rangle$ are VEVs. For $\mathfrak{h}$ in Eq.~\eqref{RotatedDoublet}, we have 
$ \langle  \mathfrak{h}_2 \rangle = v_H/\sqrt{2}, \mathfrak{h}_{1}=\chi^+, \mathfrak{h}^{\ast 1}=\chi^-,\mathfrak{h}^{\ast 2}-\mathfrak{h}_{ 2} = -\sqrt{2}i \chi^0$.

Eq.~\eqref{Hmix3} and Eq.~\eqref{RTmix} have similar construction, cancel them one by one, and use the property of $(U^\dagger D_\mu U)_1^1= -(U^\dagger D_\mu U)_2^2$, we get

\begin{align}
    \chi^+=\frac{ v_\Sigma}{v_{\rm H}} (\phi_{22}^\ast -\phi_{11}),\quad \chi^0= 0
    \label{chitriplet}
\end{align}
where  $\langle \phi_{12}\rangle = v_\Sigma/\sqrt{2}$ is VEV of triplet. Identify $\phi^\ast_{22},\phi_{11}$ as $\phi^+, -\phi^+$ in Eq.~\eqref{Phimatrix}, we see $\chi^+=2 (v_\Sigma/v_{\rm H}) \phi^+$, which is equal to the $\rho^+$ in Eq.~\eqref{rhotriplet}. So RHTE in  tensor notation and $U$ representation
is equivalent as in the former $U$ representation
except for a different form to represent doublet, triplet and Lagrangian. A major advantage is we could use $U$ representation in tensor notation to any scalar multiplets. 

\section{$U$ representation for general scalar extensions\label{sec:general}}

Before general scalar extensions, we take complex triplet and quadruplet as example to take a look at how $U$ representation is constructed.

\subsection{Complex Triplet}
The complex triplet in tensor notation is given by $\Delta_{ij}=\Delta_{(ij)}$, its hypercharge is $Y=1$, the three components corresponds to three normalized quasi-physical states as  $\Delta_{11}=\Delta^{++}$, $\Delta_{12}=\Delta^{+}/\sqrt{2}$, 
$\Delta_{22}=\Delta^{0}$.
In $U$ representation, the rotated doublet is same as in Eq.~\eqref{RotatedDoublet}. The complex triplet is   
\beq
\Delta_{ij} = U_i^l U_j^m  \phi_{lm},
\eeq
where $\phi_{lm}$s are symmetric tensors that represent three heavy states. Together with their conjugate states, they represents six new particles.  

The Lagrangian is 
\begin{align}
{\cal L}_{\rm CT}( {\rm H}_i ,\Delta_{ij})\supset
 D^\mu {\rm H}^\dagger D_\mu  {\rm H} 
+
(D_{\mu} \Delta^{\ast ij})
(D^{\mu} \Delta_{ ij})
-V({\rm H} ,\Delta ),
\end{align}
where $V({\rm H} ,\Delta )= V(\mathfrak{h} ,\Phi )$ does not include $U$ matrix. We omit its specific formula but focus on   
kinetic mixing terms to solve $\chi$ in doublet $\mathfrak{h}$ (see Eq.~\eqref{RotatedDoublet}). %The kinetic mixing terms of doublet is same as in Eq.~\eqref{Hmix}. For 
The kinetic mixing term from 
complex triplet is 
\bea
\mathcal{L}^\text{mix}_{\Delta}&=& 
2 \langle\phi_{22}\rangle \Bigl( (U^\dagger D_\mu U)_1^{2} D^\mu \phi^{\ast 12} - (U^\dagger D_\mu U)_2^{1} D^\mu \phi_{12}  \\ & &+
(U^\dagger D_\mu U)_2^{2} (D^\mu \phi^{\ast 22}-
D^\mu \phi_{ 22})
\Bigr),
\label{Qmix}
\eea
Comparing it with Eq.~\eqref{Hmix3}, and denote $\langle\phi_{22}\rangle=\langle\phi^{\ast 22}\rangle= v_{\Delta}/\sqrt{2}$,  
$\phi_{12}=\phi^+/\sqrt{2}$, $\operatorname{Im}(\phi_{22})=\eta_\Delta/\sqrt{2}$. We get 
\beq
\chi^+ = -\frac{2 v_{\Delta}}{v_{\rm H}} \phi_{12}=  -\frac{\sqrt{2} v_{\Delta}}{v_{\rm H}} \phi^+,\quad \chi^0=-\frac{2 v _\Delta}{v_{\rm H}} \eta_\Delta
%\operatorname{Im}(\phi_{22}),
\eeq
It is interesting to see the proportional factor $\frac{\sqrt{2} v_{\Delta}}{v_{\rm H}}, \frac{2 v_\Delta}{v_{\rm H}}$ are equal to tangent of rotation angles of charged scalars and neutral scalars respectively~\cite{Du:2018eaw}.

%So only scalars that are perpendicular to Goldstones appear in doublet $\mathfrak{h}$,
%the charged Higgs $\phi^{++},  with $Q=2$ are absent.

\subsection{Quadruplet  with  $Y=3/2$}
The quadruplet in tensor notation is given by $\Theta_{ijk}=\Theta_{(ijk)}$, take the case of $Y=3/2$, the four components corresponds to four normalized quasi-physical states as  $\Theta_{111}=\Theta^{3+}$, $\Theta_{112}=\Theta^{++}/\sqrt{3}$, $\Theta_{122}=\Theta^{+}/\sqrt{3}$,
$\Theta_{222}=\Theta^{0}$.
In $U$ representation, the rotated doublet is same as in Eq.~\eqref{RotatedDoublet}. The quadruplet is   
\beq
\Theta_{ijk} = U_i^l U_j^m U_k^n \phi_{lmn},
\eeq
where $\phi_{lmn}$s are symmetric tensors that represent four heavy states. Together with their conjugate states, there exist eight new particles.  

The Lagrangian is 
\begin{align}
{\cal L}_{\rm Quatruplet}( {\rm H}_i ,\Theta_{ijk})\supset
 D^\mu {\rm H}^\dagger D_\mu  {\rm H} 
+
(D_{\mu} \Theta^{\ast ijk})
(D^{\mu} \Theta_{ ijk})
-V( {\rm H} ,\Theta),
%-V(\mathfrak{h} ,\Phi ),
\end{align}
From which the kinetic mixing term of quadruplet is 
\bea
\mathcal{L}^\text{mix}_{\Theta}&=& 
3 \langle\phi_{222}\rangle \Bigl( (U^\dagger D_\mu U)_1^{2} D^\mu \phi^{\ast 122} - (U^\dagger D_\mu U)_2^{1} D^\mu \phi_{122}  \\ &+&
(U^\dagger D_\mu U)_2^{2} (D^\mu \phi^{\ast 222}-
D^\mu \phi_{ 222})
\Bigr),
\label{Qmix}
\eea
where  $\langle\phi_{222}\rangle=\langle\phi^{\ast 222}\rangle= {v_{\Theta}/\sqrt{2}}$, $\phi_{122}=\phi^+/\sqrt{3}$, $\operatorname{Im}(\phi_{222})=\eta_4/\sqrt{2}$. $\eta_4$ is a pseudoscalar.
Comparing it with Eq.~\eqref{Hmix3}, we get 
\beq
\chi^+ = -\frac{3 v_{\Theta}}{v_{\rm H}} \phi_{122}=  -\frac{\sqrt{3} v_{\Theta}}{v_{\rm H}} \phi^+,\quad \chi^0=-\frac{3 v _\Theta}{v_H} %\operatorname{Im}(\phi_{222}),
\eta_4~.
\eeq
Thus only scalars that are perpendicular to Goldstones appear in doublet $\mathfrak{h}$,
the charged Higgs $\phi^{++}, \phi^{3+}$ with $Q=2,3$ are absent. 

Both quadruplet with $Y=3/2$ and complex triplet has only one pair of charged Higgs with $Q=\pm$ and one pseudo scalar, so that $\chi$ solutions are similar. When there are more than one pair of charged-one Higgs, the solutions of $\chi$ become more complicated. E.g., in scalar extension of quadruplet with $Y=1/2$.

\subsection{Quadruplet  with $Y=1/2$}

%The quadruplet in tensor notation is given by $\Theta_{ijk}=\Theta_{(ijk)}$, take the case of $Y=1/2$, the four components corresponds to four normalized quasi-physical states as  

Different from quadruplet with $Y=3/2$, when $Y=1/2$ the four components corresponds to
$\Theta_{111}=\Theta^{++}$, $\Theta_{112}=\Theta_1^{+}/\sqrt{3}$, $\Theta_{122}=\Theta^{0}/\sqrt{3}$,
$\Theta_{222}=\Theta_2^{-}$, 
where $\Theta_1^{+}/\Theta_1^{-}$  are different from $\Theta_2^{+}/\Theta_2^{-}$, so that in this extension there are two pair of charged-one Higgs.
From kinetic term 
$(D_{\mu} \Theta^{\ast ijk})
(D^{\mu} \Theta_{ ijk})$ we get the kinetic mixing term of quadruplet as

\bea
\mathcal{L}^\text{mix}_{\Theta}&=& 
3  \langle\phi_{122}\rangle\Bigl( (U^\dagger D_\mu U)_1^{2} (2 D^\mu \phi^{\ast 112}-D^\mu \phi_{ 222})
+ (U^\dagger D_\mu U)_2^{1} (D^\mu \phi^{\ast 222}  -2 D^\mu\phi_{112})
\nonumber
\\ &+&
(U^\dagger D_\mu U)_2^{2} (D^\mu \phi^{\ast  122}-
D^\mu \phi_{ 122})
\Bigr),
\label{Qmix}
\eea
where $\langle\phi_{122}\rangle=\langle\phi^{\ast 122}\rangle= v_{\Theta}/\sqrt{6}$ are non-zero VEVs. Furthermore, we have
$\phi_{222}=\phi_2^-$, $\phi_{112}=\phi_1^+/\sqrt{3}$, and $\operatorname{Im}(\phi_{122})=\eta_4/\sqrt{6}$. 
 Comparing it with Eq.~\eqref{Hmix3}, we get 
\beq
\chi^+ =\frac{\sqrt{3} v_{\Theta}}{v_{\rm H}} (\phi_2^+-2\phi_1^+/\sqrt{3}),\quad \chi^0=-\frac{v _\Theta}{v_H} \eta _4~,%\sqrt{6}\operatorname{Im}(\phi_{122}),
\eeq
where $\xi^+$ is a linear combinations of two charged Higgs.
%where $\phi^\ast_{222}=\phi_2^+$, $\phi_{112}=\phi_1^+/\sqrt{3}$, $\phi_2^+$ and $\phi_1^+$ are two different charged-one particle, $\eta_4=\sqrt{6}\operatorname{Im}(\phi_{122})$. 

%where $\phi_{122}=\phi^+/\sqrt{3}$ represnts a charged Higgs with $Q=1$, $\eta_4=\sqrt{6}\operatorname{Im}(\phi_{222})$ represents a pseudoscalar.
%So only scalars that are perpendicular to Goldstones appear in doublet $\mathfrak{h}$, the charged Higgs $\Phi^{++}, \Phi^{3+}$ with $Q=2,3$ are absent. 

\subsection{General scalar extensions}

As singlet does not break $SU(2)_L$ symmetry, polar coordinate is enough for HEFT matching. From doublet extension, $U$ representation is available no matter there exists custodial symmetry or not. 
E.g. for 2HDM, in Ref.s~\cite{Buchalla:2023hqk, Dittmaier:2022ivi} two Higgs doublets are written in bidoublet forms rotated by $U$ matrix in front of each, which is $U$ representation in our definition except for a minor difference in bidoublet forms.
Under tensor notation, $U$ representation is general for all  scalar multiplet extensions.

\iffalse

For doublet extension, i.e. 2HDM model, Higgs basis plus polar coordinate is feasible for HEFT matching \cite{Dawson:2023ebe, Dawson:2023oce}. 
Meanwhile, $U$ representation is also available. E.g., in ~Ref.s~\cite{Buchalla:2023hqk, Dittmaier:2022ivi}, two Higgs doublets are written in bidoublet forms rotated by $U$ matrix in front of each, which is $U$ representation in our definition except for a minor overall factor. 
This example also shows that $U$ representation is feasible for scalar multiplet extension without custodial symmetry, e.g. septet with $Y=2$. Under tensor notation, we could give a general $U$ representation for scalar 
multiplet extensions.

%As for higher multiplets, except for real triplet, quadruplet with $Y=3/2$, we also have 
%complex triplet, quadruplet with $Y=1/2$, septet with $Y=2$\footnote{This septet does not break custodial symmetry \cite{Hisano:2013sn}, but its $U$ representation could be similarly constructed for HEFT matching. An illuminating example is polar coordinates for 2HDM model~\cite{Buchalla:2023hqk, Dittmaier:2022ivi}, which is equivalent to a $U$ representation.}, etc as scalar extensions. 

their differences in $U$ representation mainly exist in $\mathfrak{h}$ solution. 

\fi

Following the above discussion, we write the multiple in the tensor form, 
\begin{align}
    \Phi_{ijklm\cdots}=U^i_{i_1}U^j_{j_1}U^k_{k_1}U^l_{l_1}U^k_{k_1}U^m_{m_1}\cdots\phi_{i_1 j_1 k_1 l_1 m_1\cdots} \label{eq:TF_2j_M}
\end{align}
the kinetic term of $\Phi$ can now be written as~\footnote{For simplicity, we will not write down the Lorentz indices explicitly and assume that the covariant derivative operator $D$ only acts on the field immediately to its right.}
\begin{align}
    &(D\Phi^{*i_1 i_2 i_3 i_4 i_5\cdots})(D\Phi_{i_1 i_2 i_3 i_4 i_5\cdots}) \nonumber \\
    =&(D\phi^{*i_1 i_2 i_3 i_4 i_5\cdots})(D\phi_{i_1 i_2 i_3 i_4 i_5\cdots})+(D{U^*}^{i_n}_{k_n}DU^{j_n}_{i_n})\phi^{*\cdots i_{n-1}k_n i_{n+1}\cdots}\phi_{\cdots i_{n-1}j_n i_{n+1}\cdots} \nonumber \\
    &+({U^*}^{i_n}_{k_n}DU^{j_n}_{i_n}D\phi^{*\cdots i_{n-1}k_n i_{n+1}\cdots}\phi_{\cdots i_{n-1}j_n i_{n+1}\cdots}+D{U^*}^{i_n}_{k_n}U^{j_n}_{i_n}\phi^{*\cdots i_{n-1}k_n i_{n+1}\cdots}D\phi_{\cdots i_{n-1}j_n i_{n+1}\cdots}) \nonumber \\
    &+({U^*}^{i_m}_{k_m}DU^{j_m}_{i_m}D{U^*}^{i_n}_{k_n}U^{j_n}_{i_n}+D{U^*}^{i_m}_{k_m}U^{j_m}_{i_m}{U^*}^{i_n}_{k_n}DU^{j_n}_{i_n}) \nonumber \\
    &\qquad\phi^{*\cdots i_{m-1}k_m i_{m+1}\cdots i_{n-1}k_n i_{n+1}\cdots}\phi_{\cdots i_{m-1}j_m i_{m+1}\cdots i_{n-1}j_n i_{n+1}\cdots}
\end{align}
where only the third term potentially describes the mixing between Goldstones and scalars once the neutral component of the multiplet develops a non-vanishing VEV. Therefore we only focus on this term and the field $\phi_{\cdots i_{n-1}j_n i_{n+1}}/\phi^{*\cdots i_{n-1}k_n i_{n+1}\cdots}$ without $D$ acting on must be the neutral component of the multiplet which should be set to its VEV later.

Without loss of generality, we assume that $y\geq 0$. The index of the neutral component is given as
\begin{align}
    \underbrace{1\cdots 1}_{j-y}\underbrace{2\cdots 2}_{j+y}
\end{align}
and the indices of the charged components with unit charge therefore can be
\begin{align}
    &\underbrace{1\cdots 1}_{j-y+1}\underbrace{2\cdots 2}_{j+y-1}\qquad\text{for positive charge} \\
    &\underbrace{1\cdots 1}_{j-y-1}\underbrace{2\cdots 2}_{j+y+1}\qquad\text{for negative charge}
\end{align}
We first consider the mixing in the neutral sector, which gives us
{\footnotesize
\begin{align}
    &{U^*}^{i_n}_{k_n}DU^{j_n}_{i_n}D\phi^{*\cdots i_{n-1}k_n i_{n+1}\cdots}\phi_{\cdots i_{n-1}j_n i_{n+1}\cdots}+D{U^*}^{i_n}_{k_n}U^{j_n}_{i_n}\phi^{*\cdots i_{n-1}k_n i_{n+1}\cdots}D\phi_{\cdots i_{n-1}j_n i_{n+1}\cdots} \nonumber \\
    \supset&{U^*}^{i_{j-y+1}}_{2}DU^{2}_{i_{j-y+1}}D\phi^{*\overbrace{1\cdots 1}^{j-y}2\overbrace{2\cdots 2}^{j+y-1}}\phi_{\underbrace{1\cdots 1}_{j-y}2\underbrace{2\cdots 2}_{j+y-1}}+D{U^*}^{i_{j-y+1}}_{2}U^{2}_{i_{j-y+1}}\phi^{*\overbrace{1\cdots 1}^{j-y}2\overbrace{2\cdots 2}^{j+y-1}}D\phi_{\underbrace{1\cdots 1}_{j-y}2\underbrace{2\cdots 2}_{j+y-1}} \nonumber \\
    &+{U^*}^{i_{j-y}}_{1}DU^{1}_{i_{j-y}}D\phi^{*\overbrace{1\cdots 1}^{j-y-1}1\overbrace{2\cdots 2}^{j+y}}\phi_{\underbrace{1\cdots 1}_{j-y-1}1\underbrace{2\cdots 2}_{j+y}}+D{U^*}^{i_{j-y}}_{1}U^{1}_{i_{j-y}}\phi^{*\overbrace{1\cdots 1}^{j-y-1}1\overbrace{2\cdots 2}^{j+y}}D\phi_{\underbrace{1\cdots 1}_{j-y-1}1\underbrace{2\cdots 2}_{j+y}} \\
    =&2j C^{2j-1}_{j-y}\left[(U^\dagger D U)_2^2(D\phi^{0*}/\sqrt{C^{2j}_{j-y}})(\phi^{0}/\sqrt{C^{2j}_{j-y}})+(D U^\dagger U)_2^2(D\phi^{0}/\sqrt{C^{2j}_{j-y}})(\phi^{0*}/\sqrt{C^{2j}_{j-y}})\right] \nonumber \\
    &+2j C^{2j-1}_{j-y-1}\left[(U^\dagger D U)_1^1(D\phi^{0*}/\sqrt{C^{2j}_{j+y}})(\phi^{0}/\sqrt{C^{2j}_{j+y}})+(D U^\dagger U)_1^1(D\phi^{0}/\sqrt{C^{2j}_{j+y}})(\phi^{0*}/\sqrt{C^{2j}_{j+y}})\right] \\
    =&\left[(j+y)(U^\dagger D U)_2^2+(j-y)(U^\dagger D U)_1^1\right]\left(D\phi^{0*}\phi^{0}-D\phi^{0}\phi^{0*}\right) \\
    \supset&\frac{v_\phi}{\sqrt{2}}\left[(j+y)(U^\dagger D U)_2^2+(j-y)(U^\dagger D U)_1^1\right]D\left(\phi^{0*}-\phi^{0}\right) \\
    =&-i v_\phi\left[(j+y)(U^\dagger D U)_2^2+(j-y)(U^\dagger D U)_1^1\right]D\eta=-i 2yv_\phi(U^\dagger D U)_2^2 D\eta^0
\end{align}}where in the first equality we use the definition of $\phi_{\underbrace{1\cdots 1}_{j-y}\underbrace{2\cdots 2}_{j+y}}=\phi^0/\sqrt{C^{2j}_{j-y}}$ and the factors $2j C^{2j-1}_{j-y}$ and $2j C^{2j-1}_{j-y-1}$ come from the symmetrization of the indices. To arrive at the final kinetic mixing form of the neutral sector, we use the fact that $DU^\dagger U=-U^\dagger DU$ in the second equality and also the definition of $\phi^0=(v_\phi+h^0+i\eta^0)/\sqrt{2}$ and $(U^\dagger DU)_1^1=-(U^\dagger DU)_2^2$ in the remaining steps. With such mixing term, we find the physical CP-odd state in the Higgs doublet
\begin{align}
    \chi^0=-\frac{2y v_\phi}{v_H}\eta^0
\end{align}

The kinetic mixing in the singly-charged sector can be calculated in the same way, which gives
{\footnotesize
\begin{align}
    &{U^*}^{i_n}_{k_n}DU^{j_n}_{i_n}D\phi^{*\cdots i_{n-1}k_n i_{n+1}\cdots}\phi_{\cdots i_{n-1}j_n i_{n+1}\cdots}+D{U^*}^{i_n}_{k_n}U^{j_n}_{i_n}\phi^{*\cdots i_{n-1}k_n i_{n+1}\cdots}D\phi_{\cdots i_{n-1}j_n i_{n+1}\cdots} \nonumber \\
    \supset&{U^*}^{i_{j-y+1}}_{1}DU^{2}_{i_{j-y+1}}D\phi^{*\overbrace{1\cdots 1}^{j-y}1\overbrace{2\cdots 2}^{j+y-1}}\phi_{\underbrace{1\cdots 1}_{j-y}2\underbrace{2\cdots 2}_{j+y-1}}+D{U^*}^{i_{j-y+1}}_{2}U^{1}_{i_{j-y+1}}\phi^{*\overbrace{1\cdots 1}^{j-y}2\overbrace{2\cdots 2}^{j+y-1}}D\phi_{\underbrace{1\cdots 1}_{j-y}1\underbrace{2\cdots 2}_{j+y-1}} \nonumber \\
    &+{U^*}^{i_{j-y}}_{2}DU^{1}_{i_{j-y}}D\phi^{*\overbrace{1\cdots 1}^{j-y-1}2\overbrace{2\cdots 2}^{j+y}}\phi_{\underbrace{1\cdots 1}_{j-y-1}1\underbrace{2\cdots 2}_{j+y}}+D{U^*}^{i_{j-y}}_{1}U^{2}_{i_{j-y}}\phi^{*\overbrace{1\cdots 1}^{j-y-1}1\overbrace{2\cdots 2}^{j+y}}D\phi_{\underbrace{1\cdots 1}_{j-y-1}2\underbrace{2\cdots 2}_{j+y}} \\
    =&2j C^{2j-1}_{j-y}\left[(U^\dagger D U)_1^2(D\phi^{+*}/\sqrt{C^{2j}_{j-y+1}})(\phi^0/\sqrt{C^{2j}_{j-y}})+(D U^\dagger U)_2^1(D\phi^+/\sqrt{C^{2j}_{j-y+1}})(\phi^{0*}/\sqrt{C^{2j}_{j-y}})\right] \nonumber \\
    &+2j C^{2j-1}_{j-y-1}\left[(U^\dagger D U)_2^1(D\phi^{-*}/\sqrt{C^{2j}_{j-y-1}})(\phi^0/\sqrt{C^{2j}_{j+y}})+(D U^\dagger U)_1^2(D\phi^-/\sqrt{C^{2j}_{j-y-1}})(\phi^{0*}/\sqrt{C^{2j}_{j+y}})\right] \\
    %=&\sqrt{(j+y)(j-y+1)}\left[(U^\dagger D U)_1^2 D\phi^{+*}\phi^0+(D U^\dagger U)_2^1 D\phi^+ \phi^{0*}\right] \nonumber \\
    %&+\sqrt{(j-y)(j+y+1)}\left[(U^\dagger D U)_2^1 D\phi^{-*}\phi^0+(D U^\dagger U)_1^2 D\phi^- \phi^{0*}\right] \\
    =&(U^\dagger D U)_1^2\left[\sqrt{(j+y)(j-y+1)}D\phi^{+*}\phi^0-\sqrt{(j-y)(j+y+1)}D\phi^-\phi^{0*}\right] \nonumber \\
    &\qquad+(U^\dagger D U)_2^1\left[\sqrt{(j-y)(j+y+1)}D\phi^{-*}\phi^0-\sqrt{(j+y)(j-y+1)}D\phi^+\phi^{0*}\right] \\
    \supset&v_\phi/\sqrt{2}(U^\dagger D U)_1^2\left[\sqrt{(j+y)(j-y+1)}D\phi^{+*}-\sqrt{(j-y)(j+y+1)}D\phi^-\right] \nonumber \\
    &\qquad+v_\phi/\sqrt{2}(U^\dagger D U)_2^1\left[\sqrt{(j-y)(j+y+1)}D\phi^{-*}-\sqrt{(j+y)(j-y+1)}D\phi^+\right]
\end{align}}where in the first equality we use the definition of $\phi_{\underbrace{1\cdots 1}_{j-y+1}\underbrace{2\cdots 2}_{j+y-1}}=\phi^+/\sqrt{C^{2j}_{j-y+1}}$, $\phi_{\underbrace{1\cdots 1}_{j-y-1}\underbrace{2\cdots 2}_{j+y+1}}=\phi^-/\sqrt{C^{2j}_{j-y-1}}$ and $\phi_{\underbrace{1\cdots 1}_{j-y}\underbrace{2\cdots 2}_{j+y}}=\phi^0/\sqrt{C^{2j}_{j-y}}$. Note that $\phi^+\neq(\phi^-)^*$. The factors $2j C^{2j-1}_{j-y,\, j-y-1}$ are still from the symmetrization of the indices. We use the relation $DU^\dagger U=-U^\dagger DU$ and substitute the neutral field $\phi^0$ with its VEV $v_\phi/\sqrt{2}$ to arrive at the final kinetic mixing terms between charged Goldstones and scalars, which provides us the physical singly-charged scalar in the doublet as
\begin{align}
    \chi^+=\frac{v_\phi}{v_H}(\sqrt{(j-y)(j+y+1)}\phi^{-*}-\sqrt{(j+y)(j-y+1)}\phi^+)
\end{align}

In the above discussion, we always assume that $\Phi$ is complex. For the case where $\Phi$ is real, the hypercharge $y$ must vanish and $j$ can only be integer. The tensor representation of a real $2j$-multiplet has exact same form (Eq.~\eqref{eq:TF_2j_M}) as a complex one, however, one should further impose the realness condition
\begin{align}  
    \phi^{*\overbrace{1\cdots 1}^{j-k}\overbrace{2\cdots 2}^{j+k}}=(-1)^{k}\phi_{\underbrace{1\cdots 1}_{j+k}\underbrace{2\cdots 2}_{j-k}}\qquad  \text{with } k=-j,\cdots, j   
\end{align}
Therefore the neutral component $\phi_{\underbrace{1\cdots 1}_{j}\underbrace{2\cdots 2}_{j}}$ is pure real, which indicates there is no mixing in the neutral sector and $\chi^0=0$. To be consistent with our previous result for a real triplet, we define that
\begin{align}
    \phi_{\underbrace{1\cdots 1}_{j+1}\underbrace{2\cdots 2}_{j-1}}=-\phi^+
\end{align}
which denotes the positive singly-charged scalar. The real condition implies $\phi^{*\overbrace{1\cdots 1}^{j-1}\overbrace{2\cdots 2}^{j+1}}=\phi^{+*}\equiv\phi^-$. One should note that for the real scalar there is an extra factor of $1/2$ in the kinetic term and the VEV of the neutral component is $v_\phi$ instead of $v_\phi/\sqrt{2}$, therefore we find that the kinetic mixing of the singly-charged particles is
\begin{align}
    &v_\phi\sqrt{j(j+1)}\left[(U^\dagger D U)_1^2 D\phi^- +(U^\dagger D U)_2^1 D\phi^+\right]
\end{align}
and the physical $\chi^+$ is given as
\begin{align}
    \chi^+=\frac{\sqrt{2j(j+1)v_\phi}}{v_H}\phi^+
\end{align}

\section{Conclusion and Discussion\label{sec:con}}
We give a non-linear representation of general scalar multiplet extension of standard model, denoted as $U$ representation. Its main feature is 
to reform scalar multiplet (include doublet) as Goldstone Higgs multiplying physical states. Due to Goldstone Higgs are separated out completely in a  unitary matrix, this non-linear representation is suitable for HEFT matching. It is also helpful for tadpole renormalization~\cite{Dittmaier:2022ivi}.
We show general forms for one scalar multiplet extension. The same idea is easily available for UV-complete models with two or more scalar multiplet extensions, etc. Georgi-Machacek model.

\acknowledgments
X.W. thanks Dr. Lu Yang for useful discussion about matrix calculation and Dr. Chih-Hao Fu for algebra relations. X.W. is supported by the National Science Foundation of China under Grants No. 11947416.
\bibliographystyle{JHEP}
\bibliography{reference}

\providecommand{\href}[2]{#2}\begingroup\raggedright\begin{thebibliography}{10}

\bibitem{CMS:2012qbp}
{\bf CMS} Collaboration, S.~Chatrchyan et~al., {\it {Observation of a New Boson at a Mass of 125 GeV with the CMS Experiment at the LHC}},  {\em Phys. Lett. B} {\bf 716} (2012) 30--61, [\href{http://arxiv.org/abs/1207.7235}{{\tt arXiv:1207.7235}}].

\bibitem{ATLAS:2012yve}
{\bf ATLAS} Collaboration, G.~Aad et~al., {\it {Observation of a new particle in the search for the Standard Model Higgs boson with the ATLAS detector at the LHC}},  {\em Phys. Lett. B} {\bf 716} (2012) 1--29, [\href{http://arxiv.org/abs/1207.7214}{{\tt arXiv:1207.7214}}].

\bibitem{Brivio:2017vri}
I.~Brivio and M.~Trott, {\it {The Standard Model as an Effective Field Theory}},  {\em Phys. Rept.} {\bf 793} (2019) 1--98, [\href{http://arxiv.org/abs/1706.08945}{{\tt arXiv:1706.08945}}].

\bibitem{Appelquist:1980vg}
T.~Appelquist and C.~W. Bernard, {\it {Strongly Interacting Higgs Bosons}},  {\em Phys. Rev. D} {\bf 22} (1980) 200.

\bibitem{Longhitano:1980iz}
A.~C. Longhitano, {\it {Heavy Higgs Bosons in the Weinberg-Salam Model}},  {\em Phys. Rev. D} {\bf 22} (1980) 1166.

\bibitem{Longhitano:1980tm}
A.~C. Longhitano, {\it {Low-Energy Impact of a Heavy Higgs Boson Sector}},  {\em Nucl. Phys. B} {\bf 188} (1981) 118--154.

\bibitem{Feruglio:1992wf}
F.~Feruglio, {\it {The Chiral approach to the electroweak interactions}},  {\em Int. J. Mod. Phys. A} {\bf 8} (1993) 4937--4972, [\href{http://arxiv.org/abs/hep-ph/9301281}{{\tt hep-ph/9301281}}].

\bibitem{Herrero:1993nc}
M.~J. Herrero and E.~Ruiz~Morales, {\it {The Electroweak chiral Lagrangian for the Standard Model with a heavy Higgs}},  {\em Nucl. Phys. B} {\bf 418} (1994) 431--455, [\href{http://arxiv.org/abs/hep-ph/9308276}{{\tt hep-ph/9308276}}].

\bibitem{Herrero:1994iu}
M.~J. Herrero and E.~Ruiz~Morales, {\it {Nondecoupling effects of the SM higgs boson to one loop}},  {\em Nucl. Phys. B} {\bf 437} (1995) 319--355, [\href{http://arxiv.org/abs/hep-ph/9411207}{{\tt hep-ph/9411207}}].

\bibitem{Grinstein:2007iv}
B.~Grinstein and M.~Trott, {\it {A Higgs-Higgs bound state due to new physics at a TeV}},  {\em Phys. Rev. D} {\bf 76} (2007) 073002, [\href{http://arxiv.org/abs/0704.1505}{{\tt arXiv:0704.1505}}].

\bibitem{Buchalla:2012qq}
G.~Buchalla and O.~Cata, {\it {Effective Theory of a Dynamically Broken Electroweak Standard Model at NLO}},  {\em JHEP} {\bf 07} (2012) 101, [\href{http://arxiv.org/abs/1203.6510}{{\tt arXiv:1203.6510}}].

\bibitem{Alonso:2012px}
R.~Alonso, M.~B. Gavela, L.~Merlo, S.~Rigolin, and J.~Yepes, {\it {The Effective Chiral Lagrangian for a Light Dynamical ''Higgs Particle''}},  {\em Phys. Lett. B} {\bf 722} (2013) 330--335, [\href{http://arxiv.org/abs/1212.3305}{{\tt arXiv:1212.3305}}]. [Erratum: Phys.Lett.B 726, 926 (2013)].

\bibitem{Buchalla:2013rka}
G.~Buchalla, O.~Cat\`a, and C.~Krause, {\it {Complete Electroweak Chiral Lagrangian with a Light Higgs at NLO}},  {\em Nucl. Phys. B} {\bf 880} (2014) 552--573, [\href{http://arxiv.org/abs/1307.5017}{{\tt arXiv:1307.5017}}]. [Erratum: Nucl.Phys.B 913, 475--478 (2016)].

\bibitem{Brivio:2013pma}
I.~Brivio, T.~Corbett, O.~J.~P. \'Eboli, M.~B. Gavela, J.~Gonzalez-Fraile, M.~C. Gonzalez-Garcia, L.~Merlo, and S.~Rigolin, {\it {Disentangling a dynamical Higgs}},  {\em JHEP} {\bf 03} (2014) 024, [\href{http://arxiv.org/abs/1311.1823}{{\tt arXiv:1311.1823}}].

\bibitem{Buchalla:2013eza}
G.~Buchalla, O.~Cat\'a, and C.~Krause, {\it {On the Power Counting in Effective Field Theories}},  {\em Phys. Lett. B} {\bf 731} (2014) 80--86, [\href{http://arxiv.org/abs/1312.5624}{{\tt arXiv:1312.5624}}].

\bibitem{Gavela:2014vra}
M.~B. Gavela, J.~Gonzalez-Fraile, M.~C. Gonzalez-Garcia, L.~Merlo, S.~Rigolin, and J.~Yepes, {\it {CP violation with a dynamical Higgs}},  {\em JHEP} {\bf 10} (2014) 044, [\href{http://arxiv.org/abs/1406.6367}{{\tt arXiv:1406.6367}}].

\bibitem{Pich:2015kwa}
A.~Pich, I.~Rosell, J.~Santos, and J.~J. Sanz-Cillero, {\it {Low-energy signals of strongly-coupled electroweak symmetry-breaking scenarios}},  {\em Phys. Rev. D} {\bf 93} (2016), no.~5 055041, [\href{http://arxiv.org/abs/1510.03114}{{\tt arXiv:1510.03114}}].

\bibitem{Alonso:2015fsp}
R.~Alonso, E.~E. Jenkins, and A.~V. Manohar, {\it {A Geometric Formulation of Higgs Effective Field Theory: Measuring the Curvature of Scalar Field Space}},  {\em Phys. Lett. B} {\bf 754} (2016) 335--342, [\href{http://arxiv.org/abs/1511.00724}{{\tt arXiv:1511.00724}}].

\bibitem{Brivio:2016fzo}
I.~Brivio, J.~Gonzalez-Fraile, M.~C. Gonzalez-Garcia, and L.~Merlo, {\it {The complete HEFT Lagrangian after the LHC Run I}},  {\em Eur. Phys. J. C} {\bf 76} (2016), no.~7 416, [\href{http://arxiv.org/abs/1604.06801}{{\tt arXiv:1604.06801}}].

\bibitem{Alonso:2016oah}
R.~Alonso, E.~E. Jenkins, and A.~V. Manohar, {\it {Geometry of the Scalar Sector}},  {\em JHEP} {\bf 08} (2016) 101, [\href{http://arxiv.org/abs/1605.03602}{{\tt arXiv:1605.03602}}].

\bibitem{Pich:2016lew}
A.~Pich, I.~Rosell, J.~Santos, and J.~J. Sanz-Cillero, {\it {Fingerprints of heavy scales in electroweak effective Lagrangians}},  {\em JHEP} {\bf 04} (2017) 012, [\href{http://arxiv.org/abs/1609.06659}{{\tt arXiv:1609.06659}}].

\bibitem{Merlo:2016prs}
L.~Merlo, S.~Saa, and M.~Sacrist\'an-Barbero, {\it {Baryon Non-Invariant Couplings in Higgs Effective Field Theory}},  {\em Eur. Phys. J. C} {\bf 77} (2017), no.~3 185, [\href{http://arxiv.org/abs/1612.04832}{{\tt arXiv:1612.04832}}].

\bibitem{Pich:2018ltt}
A.~Pich, {\it {Effective Field Theory with Nambu-Goldstone Modes}},  \href{http://arxiv.org/abs/1804.05664}{{\tt arXiv:1804.05664}}.

\bibitem{Krause:2018cwe}
C.~Krause, A.~Pich, I.~Rosell, J.~Santos, and J.~J. Sanz-Cillero, {\it {Colorful Imprints of Heavy States in the Electroweak Effective Theory}},  {\em JHEP} {\bf 05} (2019) 092, [\href{http://arxiv.org/abs/1810.10544}{{\tt arXiv:1810.10544}}].

\bibitem{Sun:2022ssa}
H.~Sun, M.-L. Xiao, and J.-H. Yu, {\it {Complete NLO operators in the Higgs effective field theory}},  {\em JHEP} {\bf 05} (2023) 043, [\href{http://arxiv.org/abs/2206.07722}{{\tt arXiv:2206.07722}}].

\bibitem{Sun:2022snw}
H.~Sun, M.-L. Xiao, and J.-H. Yu, {\it {Complete NNLO operator bases in Higgs effective field theory}},  {\em JHEP} {\bf 04} (2023) 086, [\href{http://arxiv.org/abs/2210.14939}{{\tt arXiv:2210.14939}}].

\bibitem{Graf:2022rco}
L.~Gr\'af, B.~Henning, X.~Lu, T.~Melia, and H.~Murayama, {\it {Hilbert series, the Higgs mechanism, and HEFT}},  {\em JHEP} {\bf 02} (2023) 064, [\href{http://arxiv.org/abs/2211.06275}{{\tt arXiv:2211.06275}}].

\bibitem{Gomez-Ambrosio:2022why}
R.~G\'omez-Ambrosio, F.~J. Llanes-Estrada, A.~Salas-Bern\'ardez, and J.~J. Sanz-Cillero, {\it {SMEFT is falsifiable through multi-Higgs measurements (even in the absence of new light particles)}},  {\em Commun. Theor. Phys.} {\bf 75} (2023), no.~9 095202, [\href{http://arxiv.org/abs/2207.09848}{{\tt arXiv:2207.09848}}].

\bibitem{Gomez-Ambrosio:2022qsi}
R.~G\'omez-Ambrosio, F.~J. Llanes-Estrada, A.~Salas-Bern\'ardez, and J.~J. Sanz-Cillero, {\it {Distinguishing electroweak EFTs with WLWL\textrightarrow{}n\texttimes{}h}},  {\em Phys. Rev. D} {\bf 106} (2022), no.~5 053004, [\href{http://arxiv.org/abs/2204.01763}{{\tt arXiv:2204.01763}}].

\bibitem{Alonso:2021rac}
R.~Alonso and M.~West, {\it {Roads to the Standard Model}},  {\em Phys. Rev. D} {\bf 105} (2022), no.~9 096028, [\href{http://arxiv.org/abs/2109.13290}{{\tt arXiv:2109.13290}}].

\bibitem{Asiain:2021lch}
I.~n. Asi\'ain, D.~Espriu, and F.~Mescia, {\it {Introducing tools to test Higgs boson interactions via WW scattering: One-loop calculations and renormalization in the Higgs effective field theory}},  {\em Phys. Rev. D} {\bf 105} (2022), no.~1 015009, [\href{http://arxiv.org/abs/2109.02673}{{\tt arXiv:2109.02673}}].

\bibitem{Cohen:2021ucp}
T.~Cohen, N.~Craig, X.~Lu, and D.~Sutherland, {\it {Unitarity violation and the geometry of Higgs EFTs}},  {\em JHEP} {\bf 12} (2021) 003, [\href{http://arxiv.org/abs/2108.03240}{{\tt arXiv:2108.03240}}].

\bibitem{Herrero:2021iqt}
M.~J. Herrero and R.~A. Morales, {\it {One-loop renormalization of vector boson scattering with the electroweak chiral Lagrangian in covariant gauges}},  {\em Phys. Rev. D} {\bf 104} (2021), no.~7 075013, [\href{http://arxiv.org/abs/2107.07890}{{\tt arXiv:2107.07890}}].

\bibitem{Herrero:2022krh}
M.~J. Herrero and R.~A. Morales, {\it {One-loop corrections for WW to HH in Higgs EFT with the electroweak chiral Lagrangian}},  {\em Phys. Rev. D} {\bf 106} (2022), no.~7 073008, [\href{http://arxiv.org/abs/2208.05900}{{\tt arXiv:2208.05900}}].

\bibitem{Henning:2014wua}
B.~Henning, X.~Lu, and H.~Murayama, {\it {How to use the Standard Model effective field theory}},  {\em JHEP} {\bf 01} (2016) 023, [\href{http://arxiv.org/abs/1412.1837}{{\tt arXiv:1412.1837}}].

\bibitem{Drozd:2015kva}
A.~Drozd, J.~Ellis, J.~Quevillon, and T.~You, {\it {Comparing EFT and Exact One-Loop Analyses of Non-Degenerate Stops}},  {\em JHEP} {\bf 06} (2015) 028, [\href{http://arxiv.org/abs/1504.02409}{{\tt arXiv:1504.02409}}].

\bibitem{Chiang:2015ura}
C.-W. Chiang and R.~Huo, {\it {Standard Model Effective Field Theory: Integrating out a Generic Scalar}},  {\em JHEP} {\bf 09} (2015) 152, [\href{http://arxiv.org/abs/1505.06334}{{\tt arXiv:1505.06334}}].

\bibitem{Huo:2015exa}
R.~Huo, {\it {Standard Model Effective Field Theory: Integrating out Vector-Like Fermions}},  {\em JHEP} {\bf 09} (2015) 037, [\href{http://arxiv.org/abs/1506.00840}{{\tt arXiv:1506.00840}}].

\bibitem{Huo:2015nka}
R.~Huo, {\it {Effective Field Theory of Integrating out Sfermions in the MSSM: Complete One-Loop Analysis}},  {\em Phys. Rev. D} {\bf 97} (2018), no.~7 075013, [\href{http://arxiv.org/abs/1509.05942}{{\tt arXiv:1509.05942}}].

\bibitem{Brehmer:2015rna}
J.~Brehmer, A.~Freitas, D.~Lopez-Val, and T.~Plehn, {\it {Pushing Higgs Effective Theory to its Limits}},  {\em Phys. Rev. D} {\bf 93} (2016), no.~7 075014, [\href{http://arxiv.org/abs/1510.03443}{{\tt arXiv:1510.03443}}].

\bibitem{Crivellin:2016ihg}
A.~Crivellin, M.~Ghezzi, and M.~Procura, {\it {Effective Field Theory with Two Higgs Doublets}},  {\em JHEP} {\bf 09} (2016) 160, [\href{http://arxiv.org/abs/1608.00975}{{\tt arXiv:1608.00975}}].

\bibitem{Belusca-Maito:2016dqe}
H.~B\'elusca-Ma\"\i{}to, A.~Falkowski, D.~Fontes, J.~C. Rom\~ao, and J.~a.~P. Silva, {\it {Higgs EFT for 2HDM and beyond}},  {\em Eur. Phys. J. C} {\bf 77} (2017), no.~3 176, [\href{http://arxiv.org/abs/1611.01112}{{\tt arXiv:1611.01112}}].

\bibitem{Dawson:2017vgm}
S.~Dawson and C.~W. Murphy, {\it {Standard Model EFT and Extended Scalar Sectors}},  {\em Phys. Rev. D} {\bf 96} (2017), no.~1 015041, [\href{http://arxiv.org/abs/1704.07851}{{\tt arXiv:1704.07851}}].

\bibitem{Corbett:2017ieo}
T.~Corbett, A.~Joglekar, H.-L. Li, and J.-H. Yu, {\it {Exploring Extended Scalar Sectors with Di-Higgs Signals: A Higgs EFT Perspective}},  {\em JHEP} {\bf 05} (2018) 061, [\href{http://arxiv.org/abs/1705.02551}{{\tt arXiv:1705.02551}}].

\bibitem{deBlas:2017xtg}
J.~de~Blas, J.~C. Criado, M.~Perez-Victoria, and J.~Santiago, {\it {Effective description of general extensions of the Standard Model: the complete tree-level dictionary}},  {\em JHEP} {\bf 03} (2018) 109, [\href{http://arxiv.org/abs/1711.10391}{{\tt arXiv:1711.10391}}].

\bibitem{Han:2017cfr}
H.~Han, R.~Huo, M.~Jiang, and J.~Shu, {\it {Standard Model Effective Field Theory: Integrating out Neutralinos and Charginos in the MSSM}},  {\em Phys. Rev. D} {\bf 97} (2018), no.~9 095003, [\href{http://arxiv.org/abs/1712.07825}{{\tt arXiv:1712.07825}}].

\bibitem{Jiang:2018pbd}
M.~Jiang, N.~Craig, Y.-Y. Li, and D.~Sutherland, {\it {Complete one-loop matching for a singlet scalar in the Standard Model EFT}},  {\em JHEP} {\bf 02} (2019) 031, [\href{http://arxiv.org/abs/1811.08878}{{\tt arXiv:1811.08878}}]. [Erratum: JHEP 01, 135 (2021)].

\bibitem{Craig:2019wmo}
N.~Craig, M.~Jiang, Y.-Y. Li, and D.~Sutherland, {\it {Loops and Trees in Generic EFTs}},  {\em JHEP} {\bf 08} (2020) 086, [\href{http://arxiv.org/abs/2001.00017}{{\tt arXiv:2001.00017}}].

\bibitem{Haisch:2020ahr}
U.~Haisch, M.~Ruhdorfer, E.~Salvioni, E.~Venturini, and A.~Weiler, {\it {Singlet night in Feynman-ville: one-loop matching of a real scalar}},  {\em JHEP} {\bf 04} (2020) 164, [\href{http://arxiv.org/abs/2003.05936}{{\tt arXiv:2003.05936}}]. [Erratum: JHEP 07, 066 (2020)].

\bibitem{Gherardi:2020det}
V.~Gherardi, D.~Marzocca, and E.~Venturini, {\it {Matching scalar leptoquarks to the SMEFT at one loop}},  {\em JHEP} {\bf 07} (2020) 225, [\href{http://arxiv.org/abs/2003.12525}{{\tt arXiv:2003.12525}}]. [Erratum: JHEP 01, 006 (2021)].

\bibitem{Dawson:2020oco}
S.~Dawson, S.~Homiller, and S.~D. Lane, {\it {Putting standard model EFT fits to work}},  {\em Phys. Rev. D} {\bf 102} (2020), no.~5 055012, [\href{http://arxiv.org/abs/2007.01296}{{\tt arXiv:2007.01296}}].

\bibitem{Marzocca:2020jze}
D.~Marzocca et~al., {\it {BSM Benchmarks for Effective Field Theories in Higgs and Electroweak Physics}},  \href{http://arxiv.org/abs/2009.01249}{{\tt arXiv:2009.01249}}.

\bibitem{Corbett:2021eux}
T.~Corbett, A.~Helset, A.~Martin, and M.~Trott, {\it {EWPD in the SMEFT to dimension eight}},  {\em JHEP} {\bf 06} (2021) 076, [\href{http://arxiv.org/abs/2102.02819}{{\tt arXiv:2102.02819}}].

\bibitem{Zhang:2021tsq}
D.~Zhang and S.~Zhou, {\it {Radiative decays of charged leptons in the seesaw effective field theory with one-loop matching}},  {\em Phys. Lett. B} {\bf 819} (2021) 136463, [\href{http://arxiv.org/abs/2102.04954}{{\tt arXiv:2102.04954}}].

\bibitem{Zhang:2021jdf}
D.~Zhang and S.~Zhou, {\it {Complete one-loop matching of the type-I seesaw model onto the Standard Model effective field theory}},  {\em JHEP} {\bf 09} (2021) 163, [\href{http://arxiv.org/abs/2107.12133}{{\tt arXiv:2107.12133}}].

\bibitem{Brivio:2021alv}
I.~Brivio, S.~Bruggisser, E.~Geoffray, W.~Killian, M.~Kr\"amer, M.~Luchmann, T.~Plehn, and B.~Summ, {\it {From models to SMEFT and back?}},  {\em SciPost Phys.} {\bf 12} (2022), no.~1 036, [\href{http://arxiv.org/abs/2108.01094}{{\tt arXiv:2108.01094}}].

\bibitem{Dedes:2021abc}
A.~Dedes and K.~Mantzaropoulos, {\it {Universal scalar leptoquark action for matching}},  {\em JHEP} {\bf 11} (2021) 166, [\href{http://arxiv.org/abs/2108.10055}{{\tt arXiv:2108.10055}}].

\bibitem{Dawson:2021xei}
S.~Dawson, S.~Homiller, and M.~Sullivan, {\it {Impact of dimension-eight SMEFT contributions: A case study}},  {\em Phys. Rev. D} {\bf 104} (2021), no.~11 115013, [\href{http://arxiv.org/abs/2110.06929}{{\tt arXiv:2110.06929}}].

\bibitem{Du:2022vso}
Y.~Du, X.-X. Li, and J.-H. Yu, {\it {Neutrino seesaw models at one-loop matching: discrimination by effective operators}},  {\em JHEP} {\bf 09} (2022) 207, [\href{http://arxiv.org/abs/2201.04646}{{\tt arXiv:2201.04646}}].

\bibitem{Li:2022ipc}
X.~Li, D.~Zhang, and S.~Zhou, {\it {One-loop matching of the type-II seesaw model onto the Standard Model effective field theory}},  {\em JHEP} {\bf 04} (2022) 038, [\href{http://arxiv.org/abs/2201.05082}{{\tt arXiv:2201.05082}}].

\bibitem{Dawson:2022cmu}
S.~Dawson, D.~Fontes, S.~Homiller, and M.~Sullivan, {\it {Role of dimension-eight operators in an EFT for the 2HDM}},  {\em Phys. Rev. D} {\bf 106} (2022), no.~5 055012, [\href{http://arxiv.org/abs/2205.01561}{{\tt arXiv:2205.01561}}].

\bibitem{Zhang:2022osj}
D.~Zhang, {\it {Complete one-loop structure of the type-(I+II) seesaw effective field theory}},  {\em JHEP} {\bf 03} (2023) 217, [\href{http://arxiv.org/abs/2208.07869}{{\tt arXiv:2208.07869}}].

\bibitem{Liao:2022cwh}
Y.~Liao and X.-D. Ma, {\it {One-loop matching of scotogenic model onto standard model effective field theory up to dimension 7}},  {\em JHEP} {\bf 12} (2022) 053, [\href{http://arxiv.org/abs/2210.04270}{{\tt arXiv:2210.04270}}].

\bibitem{Ellis:2023zim}
J.~Ellis, K.~Mimasu, and F.~Zampedri, {\it {Dimension-8 SMEFT analysis of minimal scalar field extensions of the Standard Model}},  {\em JHEP} {\bf 10} (2023) 051, [\href{http://arxiv.org/abs/2304.06663}{{\tt arXiv:2304.06663}}].

\bibitem{Dawson:2023ebe}
S.~Dawson, D.~Fontes, C.~Quezada-Calonge, and J.~J. Sanz-Cillero, {\it {Matching the 2HDM to the HEFT and the SMEFT: Decoupling and perturbativity}},  {\em Phys. Rev. D} {\bf 108} (2023), no.~5 055034, [\href{http://arxiv.org/abs/2305.07689}{{\tt arXiv:2305.07689}}].

\bibitem{Li:2023cwy}
X.-X. Li, Z.~Ren, and J.-H. Yub, {\it {Complete tree-level dictionary between simplified BSM models and SMEFT d\ensuremath{\leq}7 operators}},  {\em Phys. Rev. D} {\bf 109} (2024), no.~9 095041, [\href{http://arxiv.org/abs/2307.10380}{{\tt arXiv:2307.10380}}].

\bibitem{Li:2023ohq}
X.~Li and S.~Zhou, {\it {One-loop matching of the type-III seesaw model onto the Standard Model Effective Field Theory}},  {\em JHEP} {\bf 05} (2024) 169, [\href{http://arxiv.org/abs/2309.14702}{{\tt arXiv:2309.14702}}].

\bibitem{DasBakshi:2024krs}
S.~Das~Bakshi, S.~Dawson, D.~Fontes, and S.~Homiller, {\it {Relevance of one-loop SMEFT matching in the 2HDM}},  {\em Phys. Rev. D} {\bf 109} (2024), no.~7 075022, [\href{http://arxiv.org/abs/2401.12279}{{\tt arXiv:2401.12279}}].

\bibitem{Dawson:2024ozw}
S.~Dawson, M.~Forslund, and M.~Schnubel, {\it {SMEFT matching to Z' models at dimension eight}},  {\em Phys. Rev. D} {\bf 110} (2024), no.~1 015002, [\href{http://arxiv.org/abs/2404.01375}{{\tt arXiv:2404.01375}}].

\bibitem{Criado:2017khh}
J.~C. Criado, {\it {MatchingTools: a Python library for symbolic effective field theory calculations}},  {\em Comput. Phys. Commun.} {\bf 227} (2018) 42--50, [\href{http://arxiv.org/abs/1710.06445}{{\tt arXiv:1710.06445}}].

\bibitem{DasBakshi:2018vni}
S.~Das~Bakshi, J.~Chakrabortty, and S.~K. Patra, {\it {CoDEx: Wilson coefficient calculator connecting SMEFT to UV theory}},  {\em Eur. Phys. J. C} {\bf 79} (2019), no.~1 21, [\href{http://arxiv.org/abs/1808.04403}{{\tt arXiv:1808.04403}}].

\bibitem{Fuentes-Martin:2022jrf}
J.~Fuentes-Mart\'\i{}n, M.~K\"onig, J.~Pag\`es, A.~E. Thomsen, and F.~Wilsch, {\it {A proof of concept for matchete: an automated tool for matching effective theories}},  {\em Eur. Phys. J. C} {\bf 83} (2023), no.~7 662, [\href{http://arxiv.org/abs/2212.04510}{{\tt arXiv:2212.04510}}].

\bibitem{Carmona:2021xtq}
A.~Carmona, A.~Lazopoulos, P.~Olgoso, and J.~Santiago, {\it {Matchmakereft: automated tree-level and one-loop matching}},  {\em SciPost Phys.} {\bf 12} (2022), no.~6 198, [\href{http://arxiv.org/abs/2112.10787}{{\tt arXiv:2112.10787}}].

\bibitem{Grojean:2013qca}
C.~Grojean, O.~Matsedonskyi, and G.~Panico, {\it {Light top partners and precision physics}},  {\em JHEP} {\bf 10} (2013) 160, [\href{http://arxiv.org/abs/1306.4655}{{\tt arXiv:1306.4655}}].

\bibitem{Alonso:2014wta}
R.~Alonso, I.~Brivio, B.~Gavela, L.~Merlo, and S.~Rigolin, {\it {Sigma Decomposition}},  {\em JHEP} {\bf 12} (2014) 034, [\href{http://arxiv.org/abs/1409.1589}{{\tt arXiv:1409.1589}}].

\bibitem{Hierro:2015nna}
I.~M. Hierro, L.~Merlo, and S.~Rigolin, {\it {Sigma Decomposition: The CP-Odd Lagrangian}},  {\em JHEP} {\bf 04} (2016) 016, [\href{http://arxiv.org/abs/1510.07899}{{\tt arXiv:1510.07899}}].

\bibitem{Gavela:2016vte}
M.~B. Gavela, K.~Kanshin, P.~A.~N. Machado, and S.~Saa, {\it {The linear\textendash{}non-linear frontier for the Goldstone Higgs}},  {\em Eur. Phys. J. C} {\bf 76} (2016), no.~12 690, [\href{http://arxiv.org/abs/1610.08083}{{\tt arXiv:1610.08083}}].

\bibitem{Qi:2019ocx}
Y.-H. Qi, J.-H. Yu, and S.-H. Zhu, {\it {Effective field theory perspective on next-to-minimal composite Higgs models}},  {\em Phys. Rev. D} {\bf 103} (2021), no.~1 015013, [\href{http://arxiv.org/abs/1912.13058}{{\tt arXiv:1912.13058}}].

\bibitem{Lindner:2022kxm}
A.~Lindner and K.~F. Muzakka, {\it {Matching to Higgs-Compositeness and Renormalization of the Higgs-Electroweak Chiral Lagrangian extended by a Scalar Singlet}},  \href{http://arxiv.org/abs/2201.05122}{{\tt arXiv:2201.05122}}.

\bibitem{Buchalla:2016bse}
G.~Buchalla, O.~Cata, A.~Celis, and C.~Krause, {\it {Standard Model Extended by a Heavy Singlet: Linear vs. Nonlinear EFT}},  {\em Nucl. Phys. B} {\bf 917} (2017) 209--233, [\href{http://arxiv.org/abs/1608.03564}{{\tt arXiv:1608.03564}}].

\bibitem{Dawson:2023oce}
S.~Dawson, D.~Fontes, C.~Quezada-Calonge, and J.~J. Sanz-Cillero, {\it {Is the HEFT matching unique?}},  {\em Phys. Rev. D} {\bf 109} (2024), no.~5 055037, [\href{http://arxiv.org/abs/2311.16897}{{\tt arXiv:2311.16897}}].

\bibitem{Banta:2023prj}
I.~Banta, T.~Cohen, N.~Craig, X.~Lu, and D.~Sutherland, {\it {Effective field theory of the two Higgs doublet model}},  {\em JHEP} {\bf 06} (2023) 150, [\href{http://arxiv.org/abs/2304.09884}{{\tt arXiv:2304.09884}}].

\bibitem{Arco:2023sac}
F.~Arco, D.~Domenech, M.~J. Herrero, and R.~A. Morales, {\it {Nondecoupling effects from heavy Higgs bosons by matching 2HDM to HEFT amplitudes}},  {\em Phys. Rev. D} {\bf 108} (2023), no.~9 095013, [\href{http://arxiv.org/abs/2307.15693}{{\tt arXiv:2307.15693}}].

\bibitem{Buchalla:2023hqk}
G.~Buchalla, F.~K\"onig, C.~M\"uller-Salditt, and F.~Pandler, {\it {Two-Higgs-doublet model matched to nonlinear effective theory}},  {\em Phys. Rev. D} {\bf 110} (2024), no.~1 016015, [\href{http://arxiv.org/abs/2312.13885}{{\tt arXiv:2312.13885}}].

\bibitem{Coleman:1969sm}
S.~R. Coleman, J.~Wess, and B.~Zumino, {\it {Structure of phenomenological Lagrangians. 1.}},  {\em Phys. Rev.} {\bf 177} (1969) 2239--2247.

\bibitem{Anisha:2024xxc}
Anisha, C.~Englert, R.~Kogler, and M.~Spannowsky, {\it {Higgs boson off-shell measurements probe nonlinearities}},  {\em Phys. Rev. D} {\bf 109} (2024), no.~9 095033, [\href{http://arxiv.org/abs/2402.06746}{{\tt arXiv:2402.06746}}].

\bibitem{Cheng:2022hbo}
Y.~Cheng, X.-G. He, F.~Huang, J.~Sun, and Z.-P. Xing, {\it {Electroweak precision tests for triplet scalars}},  {\em Nucl. Phys. B} {\bf 989} (2023) 116118, [\href{http://arxiv.org/abs/2208.06760}{{\tt arXiv:2208.06760}}].

\bibitem{Du:2018eaw}
Y.~Du, A.~Dunbrack, M.~J. Ramsey-Musolf, and J.-H. Yu, {\it {Type-II Seesaw Scalar Triplet Model at a 100 TeV $pp$ Collider: Discovery and Higgs Portal Coupling Determination}},  {\em JHEP} {\bf 01} (2019) 101, [\href{http://arxiv.org/abs/1810.09450}{{\tt arXiv:1810.09450}}].

\bibitem{Dittmaier:2022ivi}
S.~Dittmaier and H.~Rzehak, {\it {Electroweak renormalization based on gauge-invariant vacuum expectation values of non-linear Higgs representations. Part II. Extended Higgs sectors}},  {\em JHEP} {\bf 08} (2022) 245, [\href{http://arxiv.org/abs/2206.01479}{{\tt arXiv:2206.01479}}].

\end{thebibliography}\endgroup

\end{document}